\documentclass[traditabstract]{aa}  

\usepackage{graphicx}
\usepackage{tikz}
\usepackage{txfonts}
\usepackage{bm}
\usepackage[pdftex=true,colorlinks=true,urlcolor=blue,citecolor=blue,draft=false]{hyperref}
\newcommand\mum{\,\mu\mathrm{m}}
\newcommand\ellH{\ell_\mathrm{H}}

\newcommand{\pD}[2]{\frac{\partial #2}{\partial #1}}

\newcommand{\ts}{\textsuperscript}

\begin{document}

   \title{Thanatology in Protoplanetary Discs}

   \subtitle{The combined influence of Ohmic, Hall, and ambipolar diffusion on dead zones}

   \author{Geoffroy Lesur\inst{1,2}
          \and
          Matthew W. Kunz\inst{3}\thanks{NASA Einstein Postdoctoral Fellow.}
          \and
          S\'ebastien Fromang\inst{4}
          }

   \institute{Univ. Grenoble Alpes, IPAG, F-38000 Grenoble, France\\
   \email{geoffroy.lesur@ujf-grenoble.fr}
   \and 
   		CNRS, IPAG, F-38000 Grenoble, France
         \and
             Department of Astrophysical Sciences, 4 Ivy Lane, Peyton Hall, Princeton University, Princeton, NJ 08544, U.~S.~A.
             \and
             Laboratoire AIM, CEA/DSM--CNRS--Universit\'e Paris 7, Irfu/Service d'Astrophysique, CEA-Saclay, 91191 Gif-sur-Yvette, France
             }

   \date{Accepted 8 Apr 2014.}

 
\abstract{Protoplanetary discs are poorly ionised due to their low temperatures and high column densities, and are therefore subject to three ``non-ideal'' magnetohydrodynamic 
effects: Ohmic dissipation, ambipolar diffusion, and the Hall effect. The existence of magnetically driven turbulence in these discs has been a central question since the discovery 
of the magnetorotational instability. Early models considered Ohmic diffusion only and led to a scenario of layered accretion, in which a magnetically ``dead'' zone in the disc midplane 
is embedded within magnetically ``active'' surface layers at distances $\sim$1--10 au from the central protostellar object. Recent work has suggested that a combination of Ohmic dissipation 
and ambipolar diffusion can render both the midplane and surface layers of the disc inactive and that torques due to magnetically driven outflows are required to explain the observed accretion rates. We reassess this picture by performing three-dimensional numerical simulations that include, for the first time, all three non-ideal MHD effects. We find that the Hall effect can generically ``revive'' dead zones 
by producing a dominant azimuthal magnetic field and a large-scale Maxwell stress throughout the midplane, provided the angular velocity and magnetic field satisfy $\bm{\Omega\cdot B} > 0$. The attendant large magnetic pressure modifies the vertical density profile and substantially increases the disc scale height beyond its hydrostatic value. Outflows are produced, 
but are not necessary to explain accretion rates $\lesssim$$10^{-7}~{\rm M}_\odot~\mathrm{yr}^{-1}$. The flow in the disc midplane is essentially laminar, 
suggesting that dust sedimentation may be efficient. These results demonstrate that, if the MRI is relevant for driving mass accretion in protoplanetary discs, one must include 
the Hall effect to obtain even qualitatively correct results.}
  

   \maketitle

%
%
\section{Introduction}

The formation and evolution of protoplanetary discs is a long-standing problem in accretion theory. Observations indicate that T Tauri stars accrete mass from their surrounding discs at a rate $\sim$
$10^{-9}$--$10^{-7}~\mathrm{M_\odot~yr}^{-1}$ on timescales $\sim$$1$--$10~{\rm Myr}$ \citep{HC98}. However, the physical mechanism driving this accretion has remained elusive for decades. Such a 
mechanism not only is necessary for understanding how the central protostellar object accretes mass at the observed rates, but also represents a crucial component in any theory for the formation of planets. 
In order to effectively accrete, angular momentum must be removed from the infalling gas and transferred to other fluid elements within the disc and/or in the surrounding medium. It can either be 
\emph{transported radially}, in which case the transport is usually modelled as a viscous process (so-called $\alpha$-disc models; \citealt{SS73}), 
or \emph{transported vertically} by large-scale magnetically driven outflows or winds \citep{BP82}. 

The rediscovery of the magnetorotational instability (MRI; \citealt{V59,C60}) by \citet{BH91,BH98} in the context of astrophysical discs vastly improved our understanding of enhanced angular-momentum transport in differentially rotating plasmas. 
This instability produces vigorous turbulence and efficiently transports angular momentum outwards in discs with an effective viscosity $\alpha \gtrsim 10^{-3}$ \citep{HGB95}. 
The MRI has since become the main contender\footnote{Hydrodynamical processes are an interesting alternative, despite serious theoretical difficulties; see \cite{T14} for a review.} 
to explain accretion in many types of astrophysical systems. However, protoplanetary discs are cold ($T\lesssim 300~K$ at 1 au), dense ($n\gtrsim 10^{13}~\mathrm{cm}^{-3}$), and 
thus poorly ionised. Their evolution is well-described  by a set of magnetohydrodynamic (MHD) equations that include three non-ideal effects \citep{NH85,WN99}: Ohmic dissipation, ambipolar diffusion, and the Hall 
effect. In a magnetised plasma composed of neutral molecules, ions, and electrons, Ohmic dissipation is caused by collisions between electrons and neutrals, 
ambipolar diffusion by collisions between ions and neutrals, and the Hall effect by the velocity difference (``drift'') between electrons and ions (or, more generally, between positively and negatively charged species). 
These non-ideal effects tend to decouple the gas from the magnetic field, casting doubt upon the relevance of the MRI to protoplanetary discs \citep{BB94,W99,KB04,D04}.

The impact of Ohmic dissipation on stratified protoplanetary discs was first studied by \citet{G96} and \citet{SM99}, and led to a model of layered accretion at distances $\sim$1--10 au from the central protostellar object: 
a quiescent magnetically ``dead'' midplane is surrounded by turbulent magnetically ``active'' surface layers located at a few disc scale heights. This layered-accretion model has been studied extensively with increasingly 
complex analytical models \citep{SMUN00,FTB02,IN06a,IN06b} and three-dimensional numerical simulations \citep{FSH00,FS03,TSD07,IN08}. Depending upon the chemical network and the presence of well-mixed 
dust grains, these models predict a vertically integrated turbulent transport roughly consistent with the observed accretion rates. However, recent work by \citet{BS13} and \citet{SB13} has shown that ambipolar diffusion 
can render the surface layers magnetically inactive. Magnetically driven outflows, which naturally occur in stratified shearing boxes including a mean vertical field \citep{SI09,M12,O12,LFO13,FLLO13,BS13a}, 
are then needed to produce accretion rates comparable to those observed.

Although the influence of the Hall effect on the linear properties of the MRI is well understood \citep{W99,BT01}, its impact on the non-linear evolution, and in particular on the dynamics of putative dead zones, has 
remained mostly speculative \citep{SS02,SW03,SW05,W07,SW08,WS12}. In a recent publication, \citet[][hereafter KL13]{KL13} discovered a new saturation mechanism for Hall-dominated magnetorotational 
turbulence, whereby the vertical magnetic field self-organises into coherent, long-lived, axisymmetric (``zonal'') structures and the rate of angular-momentum transport becomes vanishingly small. While 
interesting in itself, this result was based on a crude representation of a protoplanetary disc -- an incompressible, dust-free, unstratified shearing box without ion-neutral drift. It is natural to ask whether this 
result applies to stratified models including Ohmic dissipation and ambipolar diffusion. 

This is precisely the question we address in this paper. We perform three-dimensional numerical simulations of the MRI in a compressible, vertically stratified shearing box, taking into account Ohmic dissipation, 
ambipolar diffusion, and the Hall effect. This marks the first time that all three non-ideal magnetohydrodynamic (MHD) effects have been accounted for in a numerical simulation of the MRI. To accomplish this 
feat, we have implemented an original formulation of ambipolar diffusion and the Hall effect in the finite-volume code \textsc{Pluto}. A simplified ionisation model is used to determine vertical diffusivity profiles 
for all three non-ideal MHD effects at distances of 1, 5, and 10 au from the central protostellar object.

The paper is organised as follows. In Section \ref{sec:sb} we present the equations of non-ideal MHD in the stratified shearing-box approximation. In Sections \ref{sec:num_method} and \ref{sec:diagnostics} we  
introduce our numerical method and define the diagnostics used to quantify the efficiency of angular-momentum transport and the production of outflows. We defer to an Appendix the technical details and tests 
of our numerical implementations of the Hall effect and ambipolar diffusion. We then present the model used to determine the ionisation rate and the diffusivity tensor (\S\ref{sec:ionisation}). The results are 
presented, first with models that neglect ambipolar diffusion (\S\S\ref{sec:linHall}--\ref{sec:nlHall}) and then with all three non-ideal MHD effects treated self-consistently (\S\ref{sec:themeat}). Since our 
ionisation model does not account for the chemical and dynamical effects of dust grains, we briefly address their impact on our results in Section \ref{sec:grains}. Finally, in Section \ref{sec:conclusions} we 
provide a summary of our results and comment on their implications for the evolution of protoplanetary discs and the formation of planetesimals.

%
%
\section{\label{sec:method}Method of solution}
\subsection{\label{sec:sb}The stratified shearing-box model}

We study the evolution of a poorly ionised protoplanetary disc in the shearing-box approximation \citep{HGB95}. A local patch of the disc centred at $R_0$ with extent $\sim$$H$ (the vertical scale height) rotates with the disc at a constant angular speed $\Omega=\Omega_{\mathrm{K}}(R_0)$. The local frame is defined by ($\bm{e}_x,\bm{e}_y,\bm{e}_z$), where $\bm{e}_x$ is aligned with the 
radial direction, $\bm{e}_y$ with the azimuthal direction, and $\bm{e}_z$ with the vertical direction. The computational domain has size $(L_x,L_y,L_z)$ with $L_i \sim \mathcal{O}(H)$; we assume the 
disc to be geometrically thin, i.e.~$H\ll R_0$. This allows us to neglect curvature terms -- the so-called Hill approximation \citep{H1878}. Taking the flow to be locally isothermal, the equations of non-ideal MHD governing the evolution of the mass density $\rho$, the velocity $\bm{u}$, and the magnetic field $\bm{B}$ are, respectively, 
\begin{align}
\label{eq:cont}\partial_t\rho+\bm{\nabla\cdot }\rho\bm{u}& =0 ,\\
\label{eq:motion}\partial_t\rho\bm{u}+\bm{\nabla\cdot}\rho\bm{uu}&= - c_{\rm s}^2\bm{\nabla}\rho+\bm{J\times B} -2\rho\bm{\Omega\times u}+\rho\bm{g} ,\\
\label{eq:induct}\partial_t\bm{B}-\bm{\nabla\times}(\bm{u\times B})&=-\bm{\nabla\times}\Big( \eta_{\rm O}\bm{J}+\eta_{\rm H}\bm{J\times \bm{e}_b}-\eta_{\rm A}\bm{J\times \bm{e}_b\times \bm{e}_b}\Big) ,
\end{align}
where $c_{\rm s}$ is the isothermal sound speed, $\bm{\Omega}=\Omega \,\bm{e}_z$ is the local angular velocity, $\bm{g}=2q\Omega^2 x \,\bm{e}_x-\Omega^2z \,\bm{e}_z$ is the local 
gravitational acceleration with $q=-{\rm d}\ln\Omega/{\rm d}\ln R$ ($=3/2$ for a Keplerian disc), $\bm{e}_b=\bm{B}/|\bm{B}|$, and $\bm{J}=\bm{\nabla \times B}$. We include all three non-ideal MHD 
effects: Ohmic dissipation, the Hall effect, and ambipolar diffusion, which are characterised by their respective diffusivity coefficients $\eta_{\rm O}$, $\eta_{\rm H}$, and $\eta_{\rm A}$. These coefficients are evaluated for a simple ionisation model in Section \ref{sec:ionisation}. Note that $\eta_{\rm H}$ and $\eta_{\rm A}$ are functions of $B = |\bm{B}|$, making the induction equation nonlinear.

It should be pointed out that the use of the isothermal shearing box approximation automatically removes large scale radial gradients and vertical thermal buoyancy effects

The above equations admit a simple solution corresponding to an unperturbed Keplerian disc:
\begin{equation}
\label{eq:equilibrium} \rho = \rho_0 \exp\left( -z^2 / 2 H^2 \right), \quad \bm{u} = -q\Omega x\, \bm{e}_y,\quad {\rm and}\quad \bm{B} = B_{z0} \, \bm{e}_z , 
\end{equation}
where $H\equiv c_{\rm s}/\Omega$ and $B_{z0}$ is a constant. We take this solution as our initial equilibrium state from which the MRI may develop. We often will make use of the velocity peculiar to the equilibrium shear flow, $\bm{v}=\bm{u}+q\Omega x \,\bm{e}_y$. The strength of the mean vertical magnetic field is quantified by a modified plasma $\beta_0$,
\begin{equation}\label{eqn:beta}
\beta_0\equiv\mathrm{sign}(\Omega B_{z0}) \, \frac{2\rho_0 c_{\rm s}^2}{B_{z0}^2}.
\end{equation}
This quantity equals the usual plasma $\beta$ in the midplane times a sign function that describes the \emph{polarity} of the magnetic field threading the disc. The Hall effect impacts the disc dynamics in a polarity-dependent way \citep{W99,BT01}.

The boundary conditions are shearing-periodic in the $x$ direction and periodic in the $y$ direction. The vertical boundary conditions are:
\begin{itemize}
\item vertical hydrostatic equilibrium for $\rho$,
\item zero vertical gradient for $v_x$ and $v_y$,
\item outflow for $v_z$, and
\item vertical field for $\bm{B}$.
\end{itemize}
We have checked that the final condition can be replaced by a zero-current condition on the boundary without having any significant impact on our results.

%
%
\subsection{\label{sec:num_method}Numerical method}

We use a modified version of the finite-volume code \textsc{Pluto} \citep{M07} to numerically integrate Equations (\ref{eq:cont})--(\ref{eq:induct}). Ohmic dissipation is treated using the resistivity module included in the publicly available version of the code. We have introduced original implementations of ambipolar diffusion (Appendix \ref{app:ambi}) 
and the Hall effect (Appendix \ref{app:hall}), which will be included in a future release of the code. Our version of \textsc{Pluto} implements a standard Godunov method using second-order--accurate spatial reconstruction with a monotonized central flux limiter, a modified HLL Riemann solver (described in Appendix \ref{app:hall}), and the constrained transport method (CT) of \cite{EH88}, which maintains $\bm{\nabla\cdot B}=0$ to machine precision. 
Face-centered electromotive forces (EMFs) computed by the Riemann solver are interpolated to cell corners using arithmetic averaging. \cite{FDKM10} demonstrated that such averaging can affect MRI growth rates when combined with HLLD or Roe Riemann solvers. However, since we use an HLL solver, we do not anticipate any numerical artefacts from the arithmetic averaging. Indeed, we show in Section \ref{sec:linHall} that we recover the linear properties of the MRI. The equations, including diffusion terms, are advanced in time explicitly 
using a second-order--accurate Runge-Kutta scheme.\footnote{Super-time-stepping schemes are of no use in our case since the Hall effect is directly incorporated into 
the Riemann solver and causes one of the most severe CFL conditions on the time step.}

We choose our units such that $\Omega=c_{\rm s}=\rho_0=1$, implying $H=1$. Unless otherwise stated, the total integration time is set to
 $\tau=1000\Omega^{-1}$, which corresponds to $\approx$$160$ local orbital periods. Time averages are computed from $t=200\Omega^{-1}$ to $\tau$, in order to avoid 
transients resulting from the initial conditions. We consider a box of size $4\times 8\times 12$ and resolution $64\times 64\times 192$. 
While this resolution (16 points per $H$) is poorer than that used in most contemporary ideal-MHD simulations of the MRI, we find it to be more than enough to
capture the relevant physical processes in our simulations. This is because the large magnetic diffusivities involved in our calculations render much of the flow laminar. 
Besides, at this resolution one run including all of the non-ideal MHD effects at $R_0 = 1~{\rm au}$ requires
70\,000 core hours on the best available CPUs. Doubling this resolution would lead to an increase in the CPU time by a factor of 32, making a
systematic exploration of the parameter space impractical given our computational resources. In order to test convergence, we repeated run  
1-OHA-5 with doubled resolution for $\tau = 100\Omega^{-1}$, finding a quantitative difference of  $<$$10\%$ in the transport diagnostics between 
$80\Omega^{-1}$ and $100\Omega^{-1}$ (see Table \ref{tab:runs}). This increases our confidence in the results. Unless otherwise stated, the initial conditions 
are the equilibrium (\ref{eq:equilibrium}) to which we add a white noise on $v_x$ with a typical rms amplitude of $0.057$.
 
The evolution of the MRI in our simulations generically produces outflows, and our box continuously loses mass (and horizontal magnetic flux). In order to mimic a steady state with 
a global accretion flow replenishing the disc, we use a mass renormalisation procedure: at each time step, we multiply $\rho$ in each cell by a constant factor such that the total mass in the box 
is conserved \citep{O12}. This procedure implies that the total momentum in the box is not conserved, since the velocity field is kept constant during the renormalisation.

%
%
\subsection{\label{sec:diagnostics}Diagnostics}

We use several diagnostics to quantify the efficiency of angular-momentum transport and the production of outflows. These rely on two averaging 
procedures:
\begin{align}
\langle Q\rangle&\equiv \frac{1}{L_x L_y} \iint \mathrm{d}x\,\mathrm{d}y \, Q ,\\
\overline{Q} &\equiv \frac{1}{L_x L_y \tau} \iiint \mathrm{d}x\,\mathrm{d}y\,\mathrm{d}t\, Q .
\end{align}
The amount of transport is determined by the Reynolds stress $R_{ij}=\rho v_iv_j$ and the Maxwell stress $M_{ij}=-B_iB_j$. This allows us to define an effective (radial) transport parameter
\begin{equation}\label{eqn:alpha}
\alpha\equiv \frac{\int {\rm d}z \, ( \overline{R_{xy}+M_{xy}}  )}{c_{\rm s}^2\int {\rm d}z\, \overline{ \rho} }.
\end{equation}
Outflows are characterised by the mass-loss rate
\begin{equation}\label{eqn:mdot}
\dot{M}_{\rm outflow} \equiv\frac{\overline{\rho v_z}\vert_{z=z_{\rm t}}-\overline{\rho v_z}\vert_{z=z_{\rm b}}}{\rho_0 c_{\rm s}}
\end{equation}
and the (dimensionless) surface magnetic stress
\begin{equation}\label{eqn:tstress}
T_{yz}^S \equiv\left| \frac{ \overline{M_{yz}}}{\rho_0 c_{\rm s}^2}\right|_{z=z_{\rm t}} ,
\end{equation}
evaluated at the top ($z_{\rm t}=4.5H$) and bottom ($z_{\rm b}=-4.5H$) of the disc surface. These (arbitrary) locations are chosen to be sufficiently far from the vertical boundaries and yet far enough from the disc midplane to always be in the magnetically dominated region of the outflow (i.e.~where $\beta < 1$).

These diagnostics can be directly related to large-scale quantities such as the accretion rate $\dot{M}_{\rm acc}$. Assuming a $1~{\rm M}_\odot$ central object, smooth surface density and temperature profiles, and neglecting $\mathcal{O}(1)$ pre-factors in \citet{FLLO13},
\begin{subequations}
\begin{align}
\dot{M}_\mathrm{acc}&\simeq \Sigma \Omega H^2 \left(\alpha+\frac{R}{H}\frac{T_{yz}^S}{\sqrt{2\pi}}\right)\\
\nonumber&\simeq  10^{-8} \,\left(\frac{\Sigma}{10^3~{\rm g~cm}^{-2}}\right)\left(\frac{\epsilon}{0.1}\right)^2\left(\frac{R}{1~\mathrm{au}}\right)^{1/2}\\
\label{eq:mdot} \mbox{} & \qquad\qquad \times \left[\frac{\alpha}{10^{-3}}+\left(\frac{0.1}{\epsilon}\right)\frac{T_{yz}^S}{2.5\times 10^{-4}}\right]~{\rm M}_\odot~{\rm yr}^{-1} .
\end{align}
\end{subequations}
where $\epsilon\equiv H/R$. The above expression is made up of two contributions to the accretion rate. The first term in the square brackets leads to 
the traditional viscous $\alpha$-disc accretion rate of \cite{SS73}, while the second term accounts for the stress from a magnetically driven outflow. 
For the latter, we have taken the stress to be opposite on both sides of the disc. We show later that this is not necessarily supported by our simulations; 
the contribution of this term should thus be considered an upper bound.

%
%
\subsection{Ionisation profiles}\label{sec:ionisation}

Our disc model is the minimum mass solar nebula \citep[MMSN;][]{H81}, whose column density and temperature profiles,
\begin{align}
\Sigma(R) &= 1700\, (R/\mathrm{1~au})^{-3/2}~\mathrm{g~cm}^{-2} ,\nonumber\\
T(R)&=280\, (R/\mathrm{1~au})^{-1/2}~\mathrm{K} , \nonumber
\end{align}
we sample at $R_0 = 1$, $5$, and $10~\mathrm{au}$. To compute the ionisation profiles, we assume 
hydrostatic balance in the vertical direction and consider the following contributions to 
the ionisation rate $\zeta$:
\begin{itemize}
\item X-ray ionisation due to 3 keV photons \citep[][see their eq.~21]{IG99,BG09};
\item cosmic-ray ionisation with $\zeta_\mathrm{cr}=\zeta_{0} \exp(-\Sigma/96~\mathrm{g~cm}^{-2})~\mathrm{s}^{-1}$ \citep[e.g.][]{UN81} and $\zeta_0=10^{-16}~\mathrm{s}^{-1}$, corresponding to the cosmic-ray ionisation rate observed in the direction of $\zeta$ Persei \citep{MH03} ;
\item radioactive decay with $\zeta_\mathrm{rad}=10^{-19}~\mathrm{s}^{-1}$ \citep{UN09}.
\end{itemize}
The ionisation fraction $x_{\rm e}$ is obtained by balancing these ionisation sources with dissociative recombination 
in a metal- and dust-free environment \citep{G96,FTB02} :
\begin{equation}\label{eqn:ionisation}
x_{\rm e} = \sqrt{\frac{\zeta}{n_{\rm n}\alpha_{\rm dr}}}+x_{\mathrm{FUV}} ,
\end{equation} 
where $\alpha_{\rm dr} = 3 \times 10^{-6}~T^{-1/2}~{\rm cm}^2~{\rm s}^{-1}$ is the dissociative recombination rate coefficient for molecular ions. 

\begin{figure}
\centering
\includegraphics[width=0.9\linewidth]{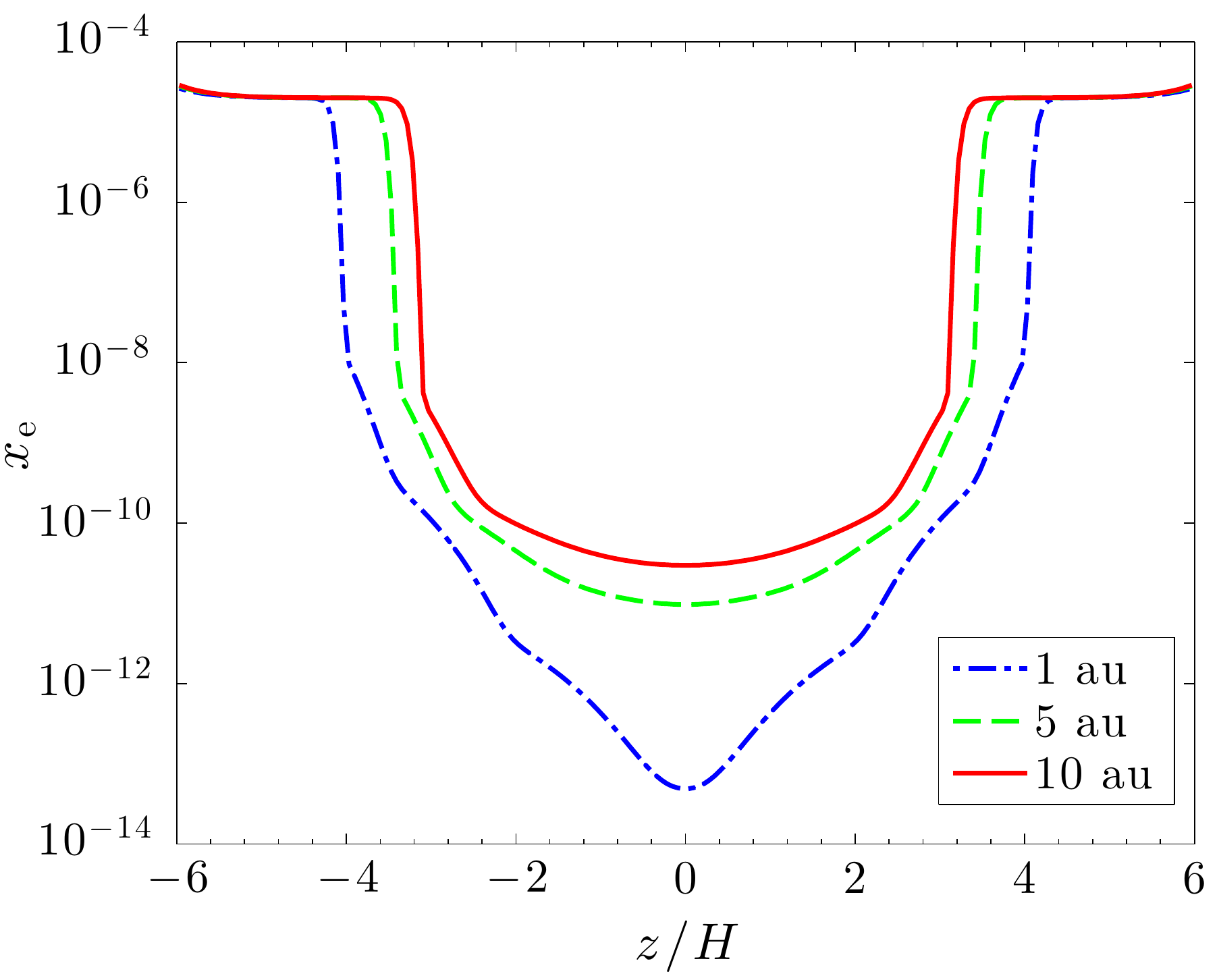}
\caption{Ionisation fraction $x_{\rm e}$ versus height $z$ at 1, 5, and 10 au.}\label{fig:ionisation_profile}
\end{figure}

The final term in Equation (\ref{eqn:ionisation}) represents the contribution from FUV radiation, 
which is known to almost fully ionise carbon and sulfur and lead to ionisation fractions $x_{\rm FUV} \sim 10^{-5}$--$10^{-4}$ for a 
penetration depth $\sim$$10^{-2}~\mathrm{g~cm}^{-2}$ \citep{PC11}. We use
\begin{equation}\label{eqn:fuv}
x_{\mathrm{FUV}}=2\times 10^{-5}\, \exp\left[-\left(\Sigma/0.03~{\rm g~cm}^{-2}\right)^4\right]
\end{equation}
as a rough estimate for the FUV ionisation fraction, which captures the essence of the FUV ionisation front. 
A similar approach has been used by \cite{BS13}, replacing the exponential in Equation (\ref{eqn:fuv}) by a step function. 
The resulting ionisation profiles at $R = 1$, $5$, and $10~{\rm au}$ are given in Fig.~\ref{fig:ionisation_profile}. For simplicity, we 
assume that these profiles are fixed throughout the course of our simulations. This simplification is not likely to be satisfied in 
actual protoplanetary disks, since changes in the density profile, as well as turbulent mixing in the disc, could lead to local changes 
in the ionisation fraction. A more sophisticated treatment, in which the chemistry evolves alongside the dynamics, will be employed in 
a future publication.

%
%

These ionisation profiles, along with a choice of $\beta_0$, determine the diffusivity coefficients for Ohmic, Hall, and ambipolar diffusion \citep[in the absence of dust grains;][]{BT01,W07}:
\begin{subequations}\label{eqn:diffusivities}
\begin{align}
\eta_\mathrm{O}&=\frac{c^2 m_{\rm e}}{4\pi e^2}\frac{n}{n_{\rm e}}\langle \sigma v\rangle_{\rm e},\\
\eta_\mathrm{H}&=\frac{Bc}{e n_{\rm e}\sqrt{4\pi}},\\
\eta_\mathrm{A}&=\frac{B^2}{\gamma_{\rm i}\rho \rho_{\rm i}},
\end{align}
\end{subequations}
where
\begin{equation}
\langle \sigma v\rangle_{\rm e} = 8.28\times 10^{-9}\,\left(\frac{T}{100~\mathrm{K}}\right)^{0.5}\,\mathrm{cm}^3~\mathrm{s} \nonumber
^{-1}
\end{equation}
is the electron-neutral collision rate \citep{D83}, $\rho_{\rm i}$ is the ion mass density, $\gamma_{\rm i}=\langle \sigma v\rangle_{\rm i}/(m_{\rm n}+m_{\rm i})$, and 
\begin{equation}
\langle \sigma v\rangle_{\rm i} = 1.3\times 10^{-9}~\mathrm{cm}^3~\mathrm{s}^{-1} \nonumber
\end{equation}
is the ion-neutral collision rate \citep{D11}; $m_{\rm n}$ and $m_{\rm i}$ are the average masses of the neutrals and ions, respectively. Due to our normalisation of the magnetic field in Equations (\ref{eq:motion}) and (\ref{eq:induct}), factors of $\sqrt{4\pi}$ and $4\pi$ appear in our expressions for the Hall and ambipolar diffusivities. 
Note that the upper layers of protoplanetary discs at 1~au are likely to be strongly ionised and heated by X-rays to temperatures up to $\sim$$8000~{\rm K}$ \citep{AK11}. 
Our simple isothermal model thus breaks down for column densities $\lesssim$$10^{-3}$--$10^{-2}~{\rm g~cm}^{-2}$. In response, we \emph{arbitrarily} multiply our diffusivities by a 
constant factor of $\exp(-\Sigma/0.01~{\rm g~cm}^{-2})$ to mimic a hot and fully ionised gas in the upper layers.

%
%
\begin{figure}
\centering
\includegraphics[width=0.9\linewidth]{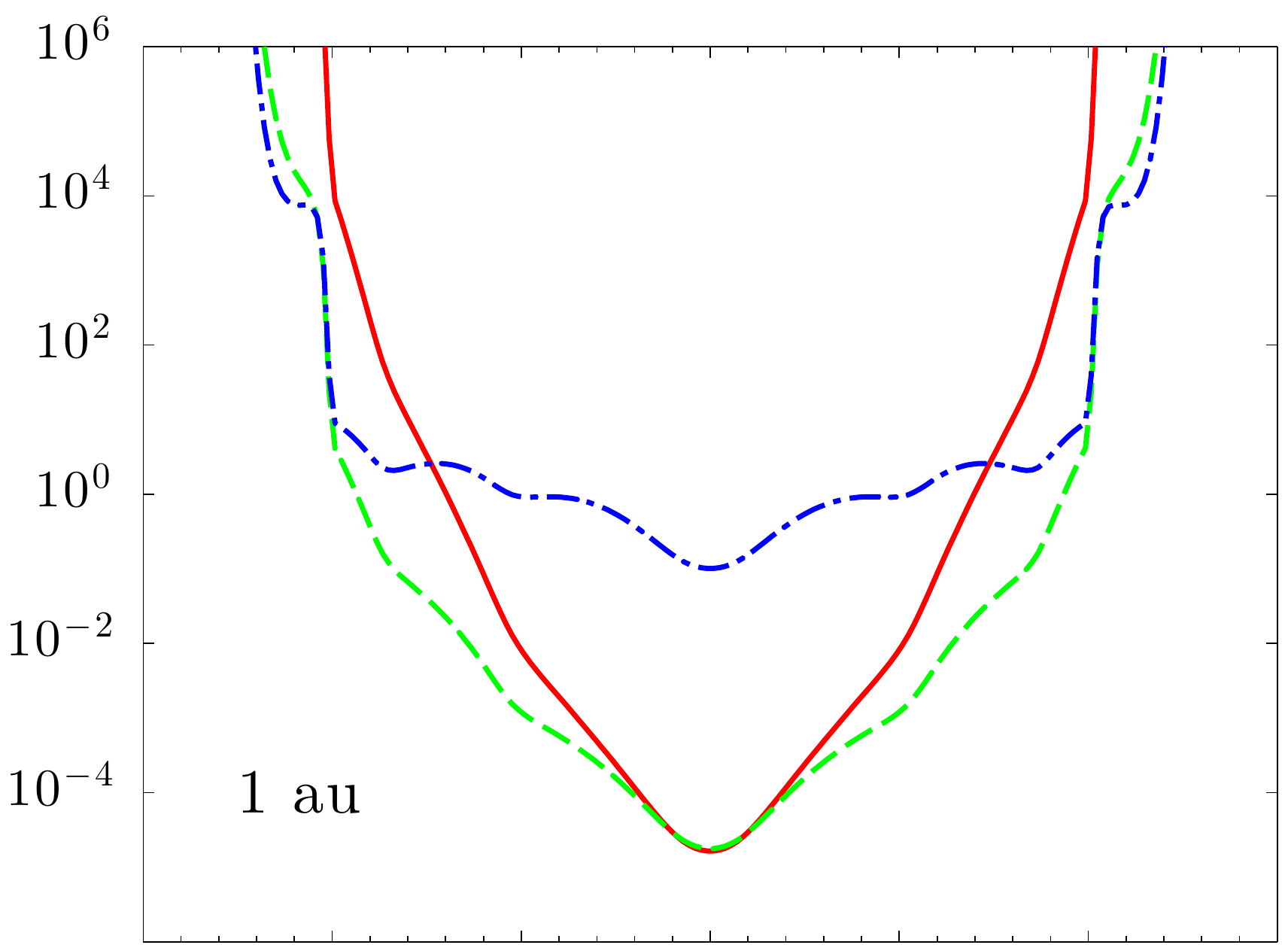}\\
\hspace{1mm}\includegraphics[width=0.91\linewidth]{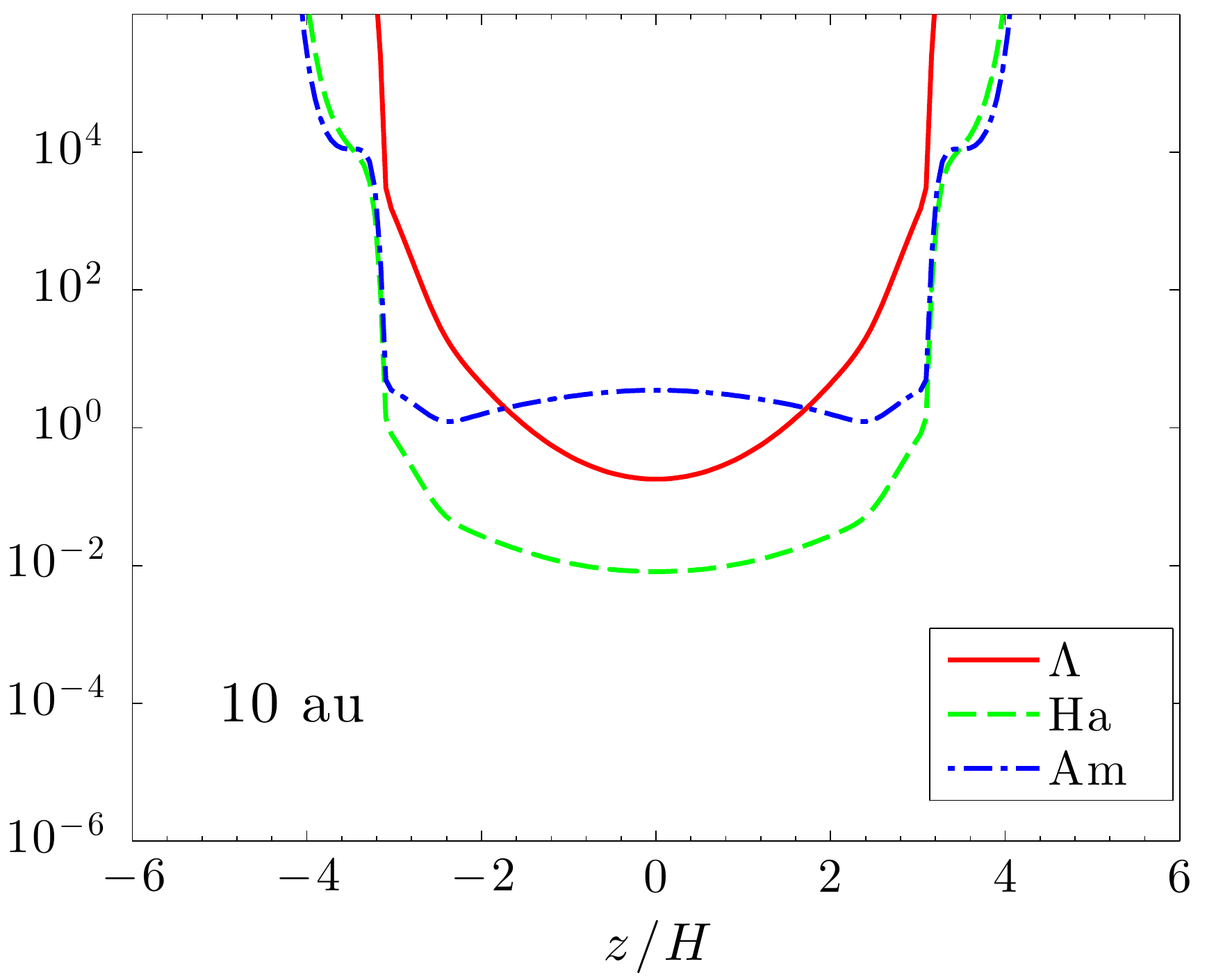}
\caption{Ohmic ($\Lambda$; eq.~\ref{eqn:lambdao}), Hall (Ha; eq.~\ref{eqn:ha}), and ambipolar (Am; eq.~\ref{eqn:am}) Elsasser numbers in the initial state versus height $z$ at 1 au (top) and 10 au 
(bottom), assuming a constant vertical magnetic field with $\beta_0=10^5$.}
\label{fig:diff_profile}
\end{figure}

%
%
\begin{figure}[t]
\centering
\includegraphics[width=0.9\linewidth]{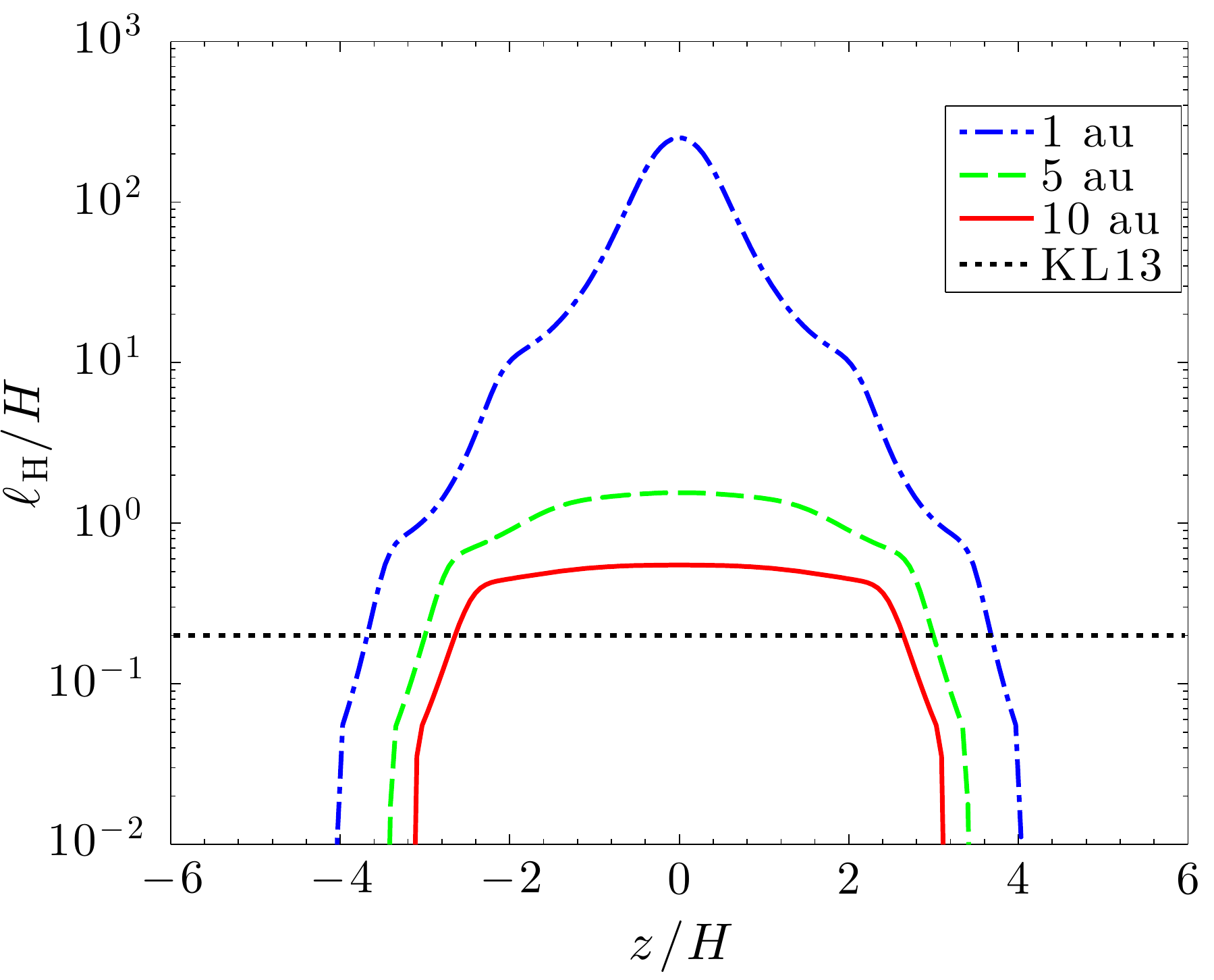}\\
\caption{Hall lengthscale $\ellH$ (eq.~\ref{eqn:ellh}) versus height $z$ at 1, 5, and 10 au. The dotted line denotes the threshold above which saturation via zonal magnetic fields occurs in unstratified simulations with net vertical magnetic flux (KL13).}
\label{fig:ellH}
\end{figure}

Introducing the Alfv\'{e}n speed $v_{\rm A} = B / \sqrt{\rho}$, the diffusivities (\ref{eqn:diffusivities}) can be cast in terms of dimensionless Elsasser numbers:
\begin{eqnarray}
\label{eqn:lambdao}\Lambda_\eta &\equiv&\frac{v_{\rm A}^2}{\Omega\eta_\mathrm{O}},\\
\label{eqn:ha}\mathrm{Ha}&\equiv&\frac{v_{\rm A}^2}{\Omega\eta_\mathrm{H}},\\
\label{eqn:am}\mathrm{Am}&\equiv&\frac{v_{\rm A}^2}{\Omega\eta_\mathrm{A}}.
\end{eqnarray}
Example profiles of these Elsasser numbers are given  
in Fig.~\ref{fig:diff_profile}; note that they evolve throughout the simulation only via changes in density and magnetic-field strength. 
These profiles are similar to those used by \citet{BS13}.\footnote{Note that \cite{BS13} 
included a numerical cap to limit diffusion coefficients in their Fig.~1 for $|z|<2$, which we do not include in our 
plot.} Another useful way of quantifying Hall diffusion -- one which is independent of the magnetic-field strength -- is the Hall lengthscale
\begin{equation}\label{eqn:ellh}
\ellH \equiv\frac{v_{\rm A}}{\Omega \mathrm{Ha}}.
\end{equation}
When $\ellH\gtrsim 0.2H$, KL13 found that unstratified simulations of the Hall-MRI with net vertical magnetic flux 
saturate by the production of strong axisymmetric (``zonal'') magnetic fields (see also Appendix \ref{app:hall}). In Fig.~\ref{fig:ellH} we present
$\ellH$ as a function of $z$ for our ionisation profiles, demonstrating that most of the disc midplane is in the putative zonal-field regime.

The non-ideal MHD terms introduce second derivatives of the magnetic field into the induction equation. This 
often yields a tight constraint on the maximum allowed stable timestep in our simulations. To circumvent this problem, we following \citet{BS13} by 
introducing a capping procedure on the magnetic diffusivities $\eta_{\mathrm{O,H,A}}$. Whenever one of these diffusivities becomes larger than $\eta_{\mathrm{cap}}=10\,\Omega H^2$, we automatically set its value to $\eta_{\mathrm{cap}}$. We have briefly explored the impact of the cap value on our results: increasing $\eta_{\rm cap}$ for $\eta_\mathrm{O}$ and $\eta_\mathrm{A}$ does not impact quantitatively our results at 1 au. However, increasing $\eta_{\rm cap}$ for $\eta_\mathrm{H}$ by a factor of 4 tends to increase $\alpha$ significantly (by $\sim$$50\%$ at 1 au). Therefore, our stress values for $\Omega B_{z0} >0$ should be understood as lower bounds when the Hall effect is included.


\begin{table*}
\centering
\setlength{\tabcolsep}{10pt}
\begin{tabular}{ l cccccccc}
\hline
Simulation & $R_0$ & Ohmic & Hall & Ambipolar & $\beta_0$ & $\alpha$ & $\dot{M}_{\rm outflow}$ & $T_{yz}^S$\\
\hline

1-O-5		&   1 au    &  $\times$     &                    &              		 &  $10^5$ &  $2.5\times 10^{-3}$ & $1.3\times 10^{-4}$ &    N/A\\    	
1-OH-5          	&   1 au    &  $\times$     &   $\times$	&              		 &  $10^5$ &  $4.5\times 10^{-1}$ & $2.0\times 10^{-2}$ &    N/A\\    	
1-OA-5          	&   1 au    &  $\times$     &			&      $\times$     &  $10^5$ &  $5.9\times 10^{-4}$ & $5.8\times 10^{-5}$ &    $1.5\times 10^{-4}$\\    	
1-OHA-5          	&   1 au    &  $\times$     &   $\times$  &      $\times$     &  $10^5$ &  $5.0\times 10^{-2}$ & $1.4\times 10^{-3}$ &    $6.5\times 10^{-4}$\\    	

\hline
1-OHA-mB	&   1 au    & $\times$     &   $\times$  &      $\times$      &  $-10^5$ &  $3.9\times 10^{-4}$ & $3.8\times 10^{-5}$ &    N/A  \\    	
1-OHA-znf	&   1 au    & $\times$     &   $\times$  &      $\times$      &  $\infty$ &  $5.4\times 10^{-6}$& $1.7\times 10^{-7}$  &    N/A\\    	
1-OA-5-e  		&   1 au    & $\times$     &			&      $\times$     &  $10^5$ &  $4.7\times 10^{-4}$ & $4.6\times 10^{-5}$ &    $1.3\times 10^{-4}$\\    	

\hline
1-O-3          	&   1 au    &  $\times$     &			&              		&  $10^3$ &  $2.4\times 10^{-1}$ & $9.3\times 10^{-3}$ &    $1.6\times 10^{-2}$\\    	
1-OH-3          	&   1 au    &  $\times$     &   $\times$  &              		&  $10^3$ &  $1.5$ & $9.7\times 10^{-2}$ &    $4.7\times 10^{-2}$\\    	
1-OA-3          	&   1 au    &  $\times$     &			&      $\times$    &  $10^3$  & $2.4\times 10^{-2}$ & $8.4\times 10^{-4}$ &    $4.3\times 10^{-3}$\\    	
1-OHA-3          	&   1 au    &  $\times$     &   $\times$  &      $\times$    &  $10^3$ &  $3.1\times 10^{-1}$ & $5.5\times 10^{-3}$ &    $1.2\times 10^{-2}$\\    	
\hline
5-OHA-5		&   5 au    	&  $\times$     &   $\times$  &      $\times$    &  $10^5$ &  $2.3\times 10^{-2}$ & $4.6\times 10^{-4}$ &    $4.0\times 10^{-4}$\\    	
10-OA-5         	&   10 au 	&  $\times$    &			 &    $\times$      &  $10^5$ &  $1.3\times 10^{-3}$ & $1.3\times 10^{-4}$ &    $2.5\times 10^{-4}$\\    	
 10-OHA-5         &   10 au 	&  $\times$     &   $\times$  &      $\times$    &  $10^5$ &  $1.0\times 10^{-2}$ & $2.4\times 10^{-4}$ &    $3.5\times 10^{-4}$\\    	
 \hline
10-OA-gr            &   10 au    &  $\times$     &   $\times$  &                           &  $10^5$ &  $7.8\times 10^{-4}$ & $8.7\times 10^{-5}$ &    $1.7\times 10^{-4}$\\    	
10-OHA-gr            &   10 au    &  $\times$     &   $\times$  &     $\times$                      &  $10^5$ &  $2.4\times 10^{-3}$ & $8.7\times 10^{-5}$ &    $1.9\times 10^{-4}$\\    	
          \hline
\end{tabular}
\vspace{3mm}
\caption{\label{tab:runs}List of the runs discussed in this paper. Each simulation is of a stratified shearing box of size $4\times 8\times 12$ located a radial distance $R_0$ from the central protostellar object. The plasma $\beta_0$ is given in the initial state (see eq.~\ref{eqn:beta}). The effective (radial) transport parameter $\alpha$, the mass-loss rate due to outflows $\dot{M}_{\rm outflow}$, and the (dimensionless) surface magnetic stress $T^S_{xy}$ are defined by Equations  (\ref{eqn:alpha}), (\ref{eqn:mdot}), and (\ref{eqn:tstress}), respectively.} 
\end{table*}

%
%
\section{Results}

%
%
\subsection{\label{sec:linHall}Linear evolution of stratified discs subject to Ohmic and Hall diffusion}

%
%
\begin{table}[]
\centering
\setlength{\tabcolsep}{20pt}
\begin{tabular}{ccr}
\hline
Mode number&$\gamma~(\Omega)$ & $\sigma$\\
\hline
\hline
1 &0.728 & $1$ \\
2 &0.727 & $-1$ \\
3& 0.660 & $-1$\\
4& 0.658 & $1$ \\
5& 0.550 & $1$\\
6& 0.531 & $-1$\\
7& 0.406 & $1$\\
8& 0.372 & $-1$\\
\hline
\end{tabular}
\vspace{3mm}
\caption{\label{tab:eigenmodes}Fastest-growing eigenmodes in run 1-OH-5 with growth rate $\gamma$ (in units of $\Omega$) and symmetry label $\sigma$ (see eq.~\ref{eq:symmetries}).}
\end{table}

%
%
\begin{figure}
\centering
\includegraphics[width=0.9\linewidth]{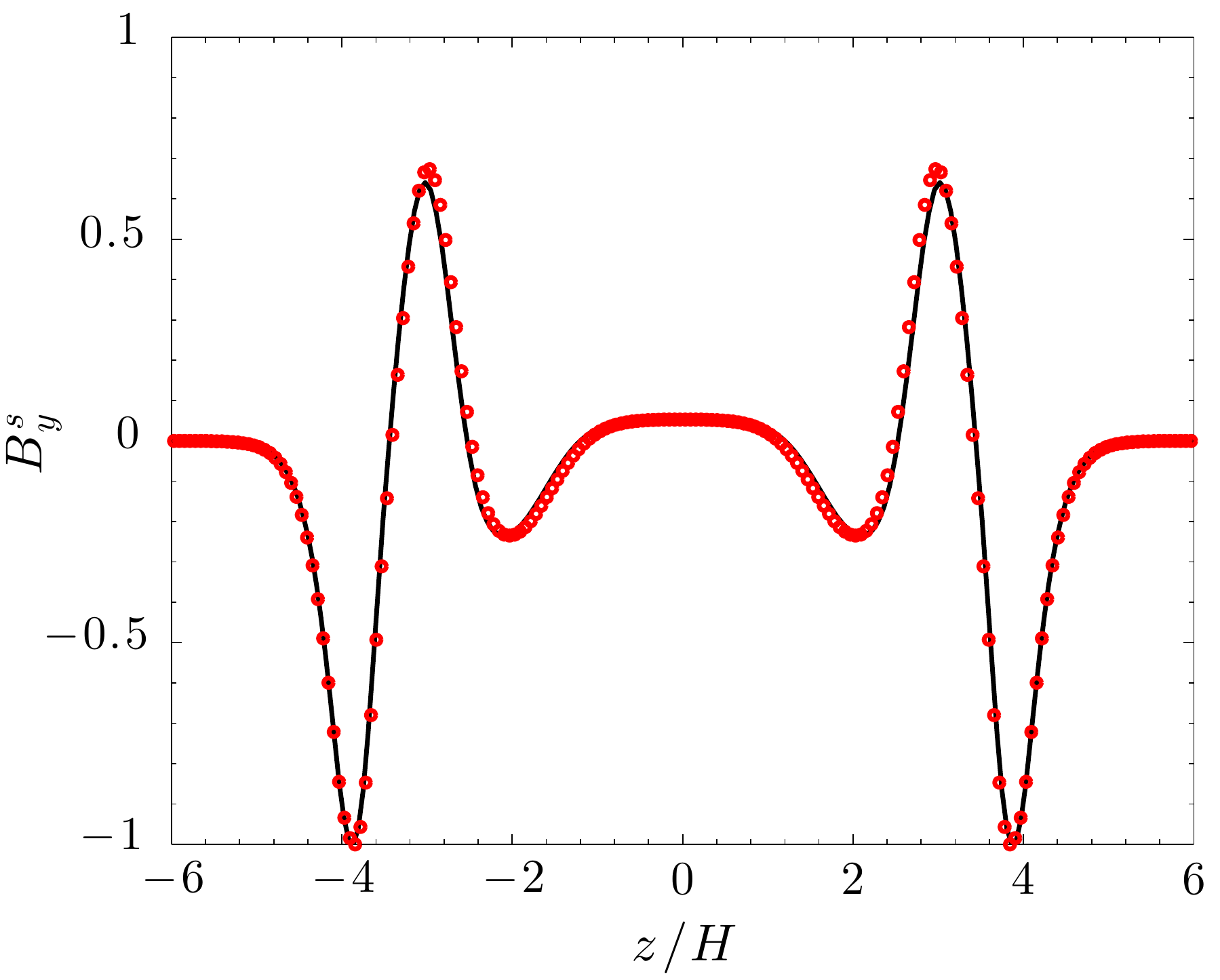}
\includegraphics[width=0.9\linewidth]{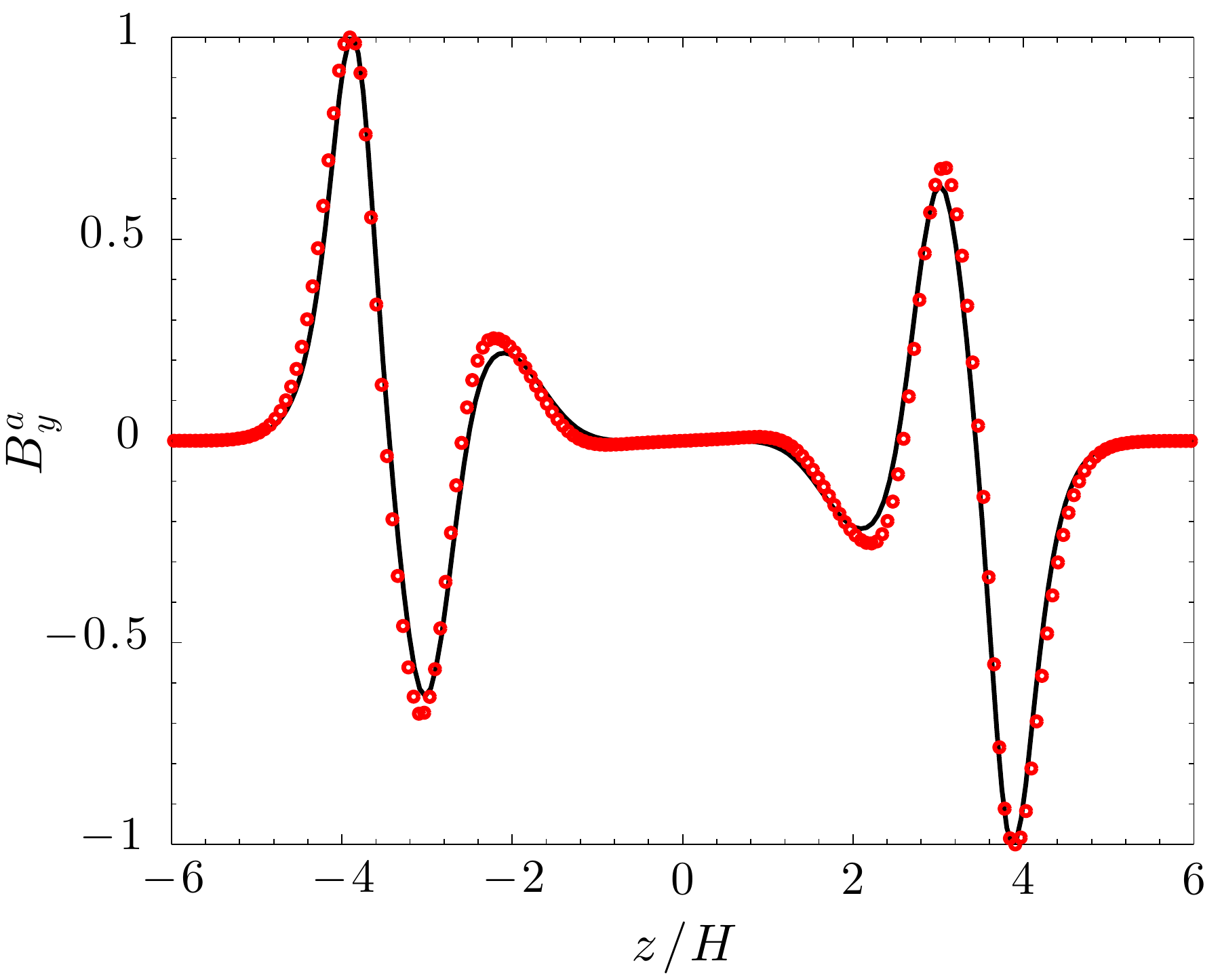}
\caption{Comparison of $B_y(z)$ from the simulation (red circles) and from the linear calculation (black line), which include Ohmic and 
Hall diffusion computed at 1~au with $\beta_0=10^5$. Top: $n=1$ eigenmode and symmetric component of $B_y(z)$. 
Bottom: $n=2$ eigenmode and antisymmetric component of $B_y(z)$.}
\label{fig:lin_profile}
\end{figure}

We first concentrate on the influence of the Hall effect on the linear evolution of an MRI-unstable, stratified disc. We set $\eta_{\rm A} = 0$, in order to isolate the effects of Hall diffusion, 
but include Ohmic dissipation. We use a setup identical to run 1-OH-5, except that we seed the instability with very small perturbations to the velocity field (RMS = $10^{-6}$) so as to produce a clean linear phase. We compare the vertical profile of the resulting azimuthal magnetic field, $B_y(z)$, at $t=10\Omega^{-1}$ with that predicted by linear theory, the latter calculated with a pseudo-spectral 
method similar to that used by \citet{LFG10}. Each linear eigenmode is characterised by its growth rate $\gamma$ and its symmetry with respect to the midplane, $\sigma=\pm 1$:
\begin{align}
B_x(-z)&=\sigma B_x(z),&v_x(-z)&=-\sigma v_x(z),\nonumber \\
B_y(-z)&=\sigma B_y(z),&v_y(-z)&=-\sigma v_y(z),\nonumber \\
B_z(-z)&=B_z(z),&v_z(-z)&=-v_z(z).\label{eq:symmetries}
\end{align}
The fastest-growing eigenmodes, their calculated growth rates and symmetries are listed in Table \ref{tab:eigenmodes}. For the ionisation profiles we consider, the $n=1$ and $n=2$ eigenmodes have very similar 
growth rates, making the identification of each individual mode difficult. To isolate each mode, we decompose $B_y(z)$ from the numerical simulation into symmetric ($\sigma=1$) and antisymmetric ($\sigma=-1$) parts, each of which can then be compared to the two fastest-growing eigenmodes obtained by linear theory (Fig.~\ref{fig:lin_profile}). We find that the most unstable modes are accurately captured by our implementation of the Hall effect at this resolution. Moreover, the measured growth rate $\gamma_{\rm num} = 0.726\,\Omega$ in our numerical simulation matches the theoretical value for the fastest growing mode to less than one per cent.

The above calculation assumed an isothermal fluid, for which buoyancy-driven modes are absent. We refer the reader to \citet{UR05} for a local, linear calculation coupling vertical buoyancy to the Hall effect.
%
%
\subsection{\label{sec:nlHall}Nonlinear evolution of stratified discs subject to Ohmic and Hall diffusion}
\subsubsection{Fiducial runs}

\begin{figure*}
\includegraphics[width=0.96\linewidth]{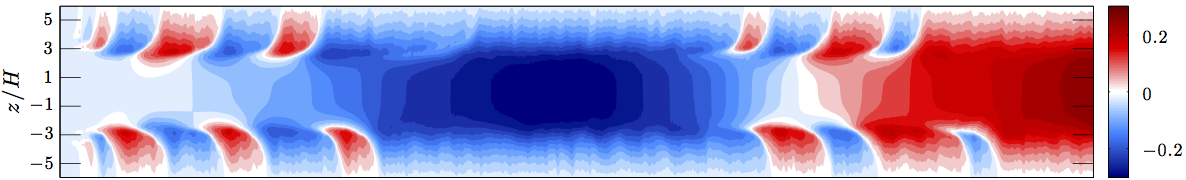}
 \includegraphics[width=0.95\linewidth]{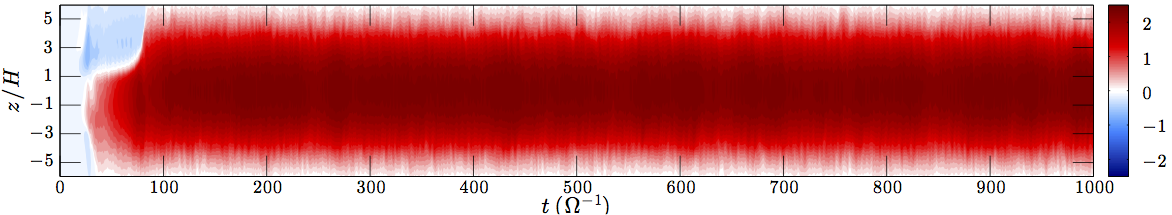}
\caption{Space-time diagram of the horizontally averaged azimuthal magnetic field, $\langle B_y \rangle$, in the Ohmic (1-O-5; top) and Ohmic-Hall (1-OH-5; bottom) runs.}
\label{fig:by_ideal}
\end{figure*}

In Fig.~\ref{fig:by_ideal} we present space-time diagrams of $\langle B_y\rangle$ for runs 1-O-5 (top) and 1-OH-5 (bottom). The Ohmic run (1-O-5) presents an alternating pattern 
in the upper layers commonly referred to as a ``butterfly diagram'', which was first described in the context of magnetorotational turbulence by 
\citet{BNST95} and \citet{SHGB96}. When the Hall effect is included (1-OH-5), this butterfly pattern is replaced by a strong quasi-steady azimuthal magnetic field that fills the 
disc midplane. This strong azimuthal field is accompanied by a weaker radial field (not shown) which feeds the azimuthal component via shear.

\begin{figure*}
\includegraphics[width=0.95\linewidth]{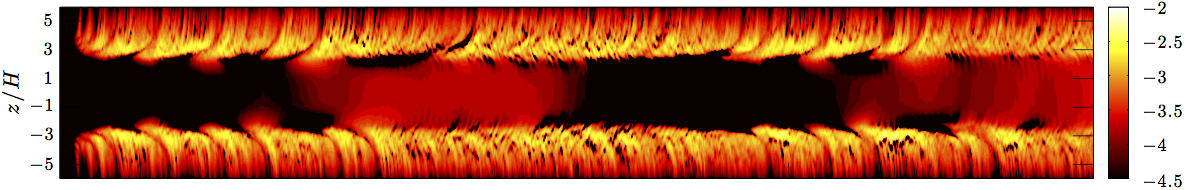}
\includegraphics[width=0.95\linewidth]{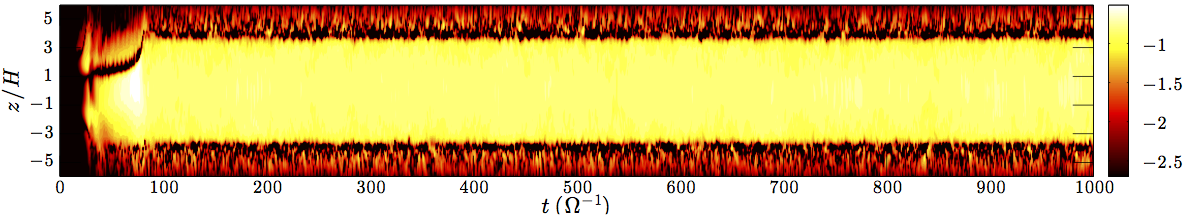}
\caption{Space-time diagram of the logarithm of the horizontally averaged Maxwell stress, $\log \langle-B_x B_y\rangle$, in the Ohmic (1-O-5; top) and Ohmic-Hall (I-OH-5; bottom) runs.}
\label{fig:ma_ideal}
\end{figure*}

The Ohmic run exhibits the usual layered accretion model \citep{G96,FS03}, with a very weak Maxwell stress in the disc midplane and a turbulent layer in the ionised atmosphere (Fig.~\ref{fig:ma_ideal}-top). 
When the Hall effect is included, we find that the magnetic stress is increased by almost two orders of magnitude (from $\alpha=2.5\times 10^{-3}$ to $\alpha=4.5\times 10^{-1}$) and is clearly located 
in the Hall-dominated region ($-3H \lesssim z \lesssim 3H$; Fig.~\ref{fig:ma_ideal}-bottom). This stress is not a ``turbulent'' stress in the usual sense, but rather is a consequence of the large-scale magnetic structure 
shown in Fig.~\ref{fig:by_ideal}. In the upper atmosphere ($|z|>4$), we find a more classical turbulent layer qualitatively similar to the active layer in run 1-O-5.

%
%

The presence of a vertical magnetic flux is known to trigger outflows in shearing-box simulations 
\citep{SI09,LFO13,FLLO13,BS13a}; our simulations exhibit outflows as well. We find that these outflows are sensitive to the
presence of the Hall effect: mass-loss rates increase by more than two orders of magnitude. However, the outflows do 
not have any well-defined orientation: they can be directed toward either positive or negative $x$, a phenomenon 
previously observed in ideal-MHD simulations presented by \cite{FLLO13} and \cite{BS13a}. This is one consequence 
of the neglect of curvature terms in the shearing-box approximation. Because of this asymmetry, we find no average stress 
$T_{yz}^S$ exerted at the disc surface by the outflow.

%
%

\begin{figure}
\centering
\includegraphics[width=0.9\linewidth]{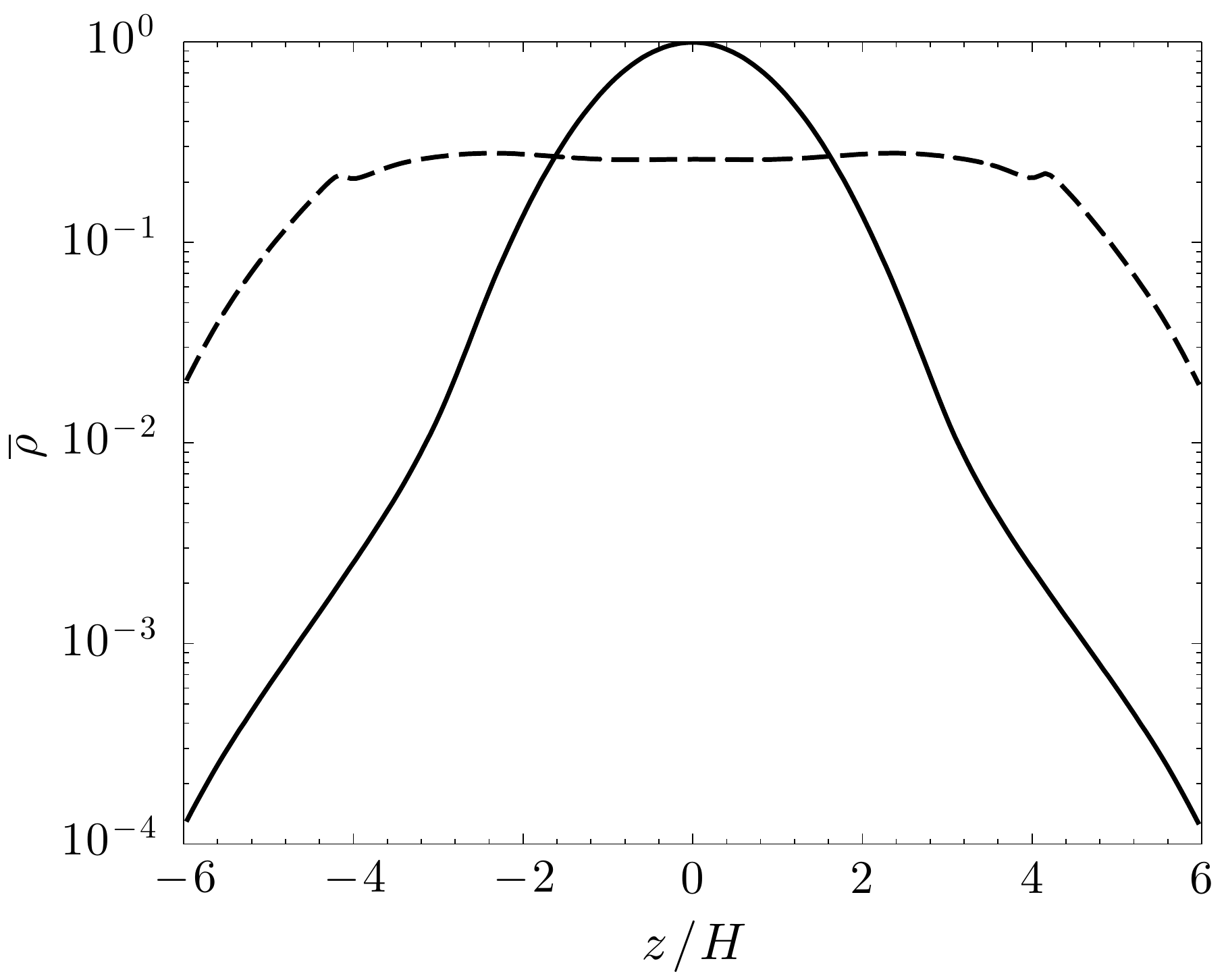}
\includegraphics[width=0.9\linewidth]{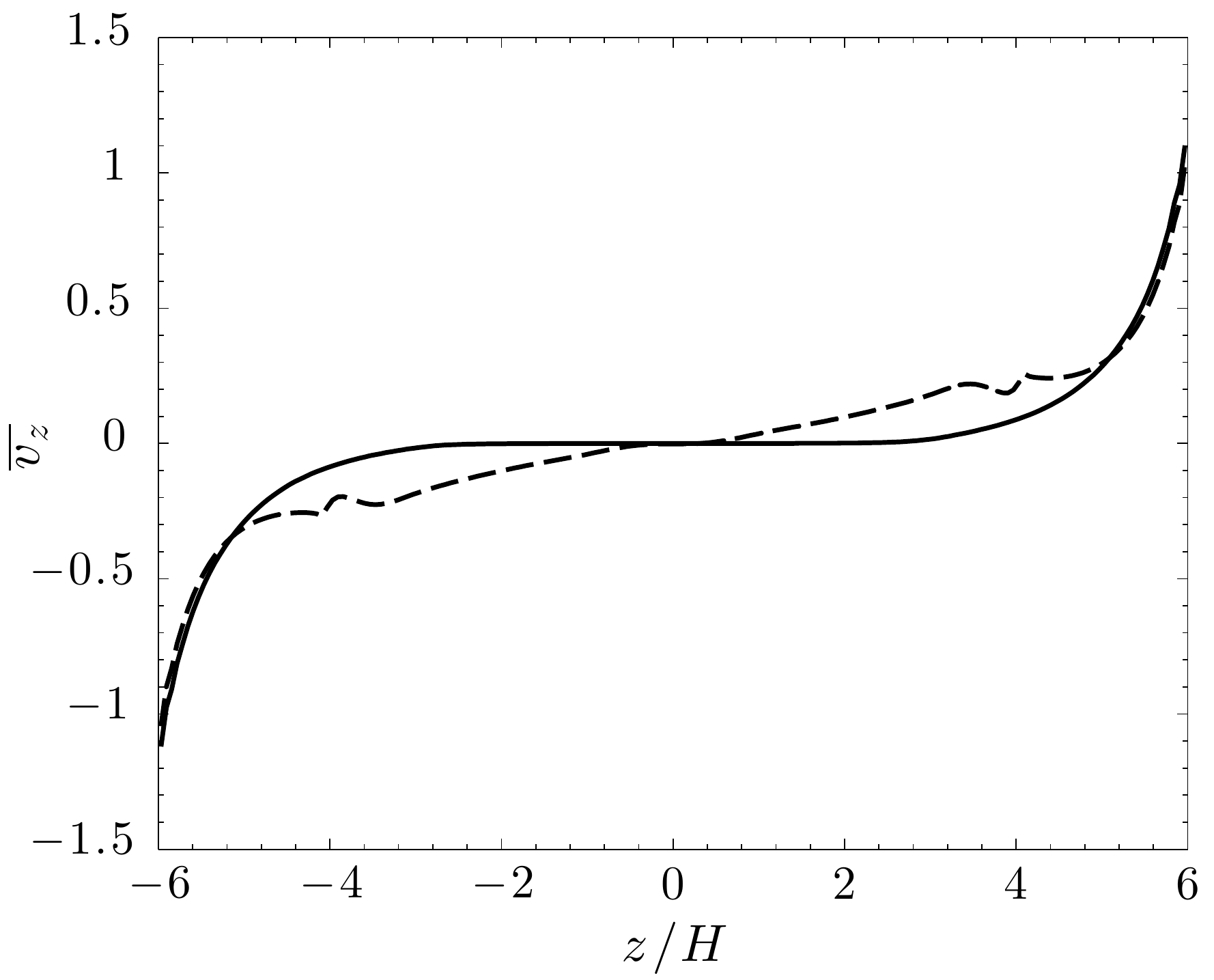}
\caption{Horizontally averaged density profile $\overline{\rho}$ (top) and vertical velocity $\overline{v_z}$ (bottom) versus height $z$, averaged over $\approx$$140$ orbits, 
from runs 1-O-5 (solid line) and 1-OH-5 (dashed line).}
\label{fig:rho_vz}
\end{figure}

The very large mass-loss rate found in the Ohmic-Hall run can be explained by the modification of the vertical 
equilibrium, which is \emph{no longer hydrostatic}. As revealed by Fig.~\ref{fig:rho_vz}, the pressure associated with the strong azimuthal  
magnetic field puffs up the disc, thereby increasing the amount of mass in the 
disc atmosphere where the outflow originates (near $z=4.5H$). On the other hand, the vertical velocity is not significantly  
affected by the Hall term. This explains the two orders of magnitude increase in the rate of mass loss between runs 1-O-5 and 1-OH-5. 

We note that a similar thickening of the disc atmosphere has been found by \cite{HT11} using resistive MHD simulations including radiative transfer. This thickening 
has been used to explain the infrared excess in Herbig stars, which cannot be explained by a disc in vertical hydrostatic equilibrium \citep{TB14}. 
Our model presents an alternative scenario, in which the strong magnetic pressure driven by the Hall-amplified azimuthal magnetic field dominates the total pressure 
in the disc midplane.

%
%

\subsubsection{Origin of the Hall-driven azimuthal field}
\label{sec:Hall-induction}

%
%
\begin{figure}[t!]
\centering
\includegraphics[width=0.9\linewidth]{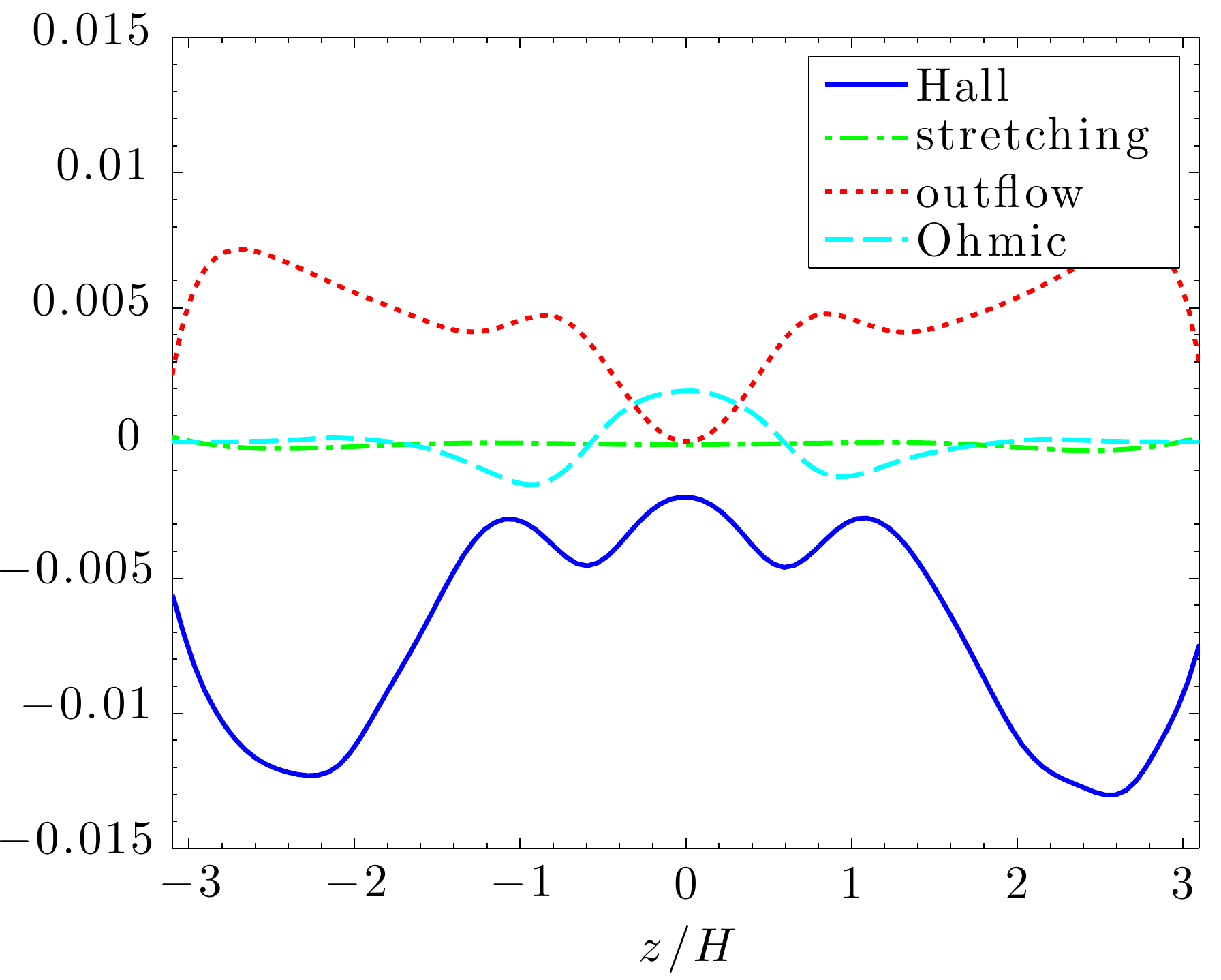}
\includegraphics[width=0.9\linewidth]{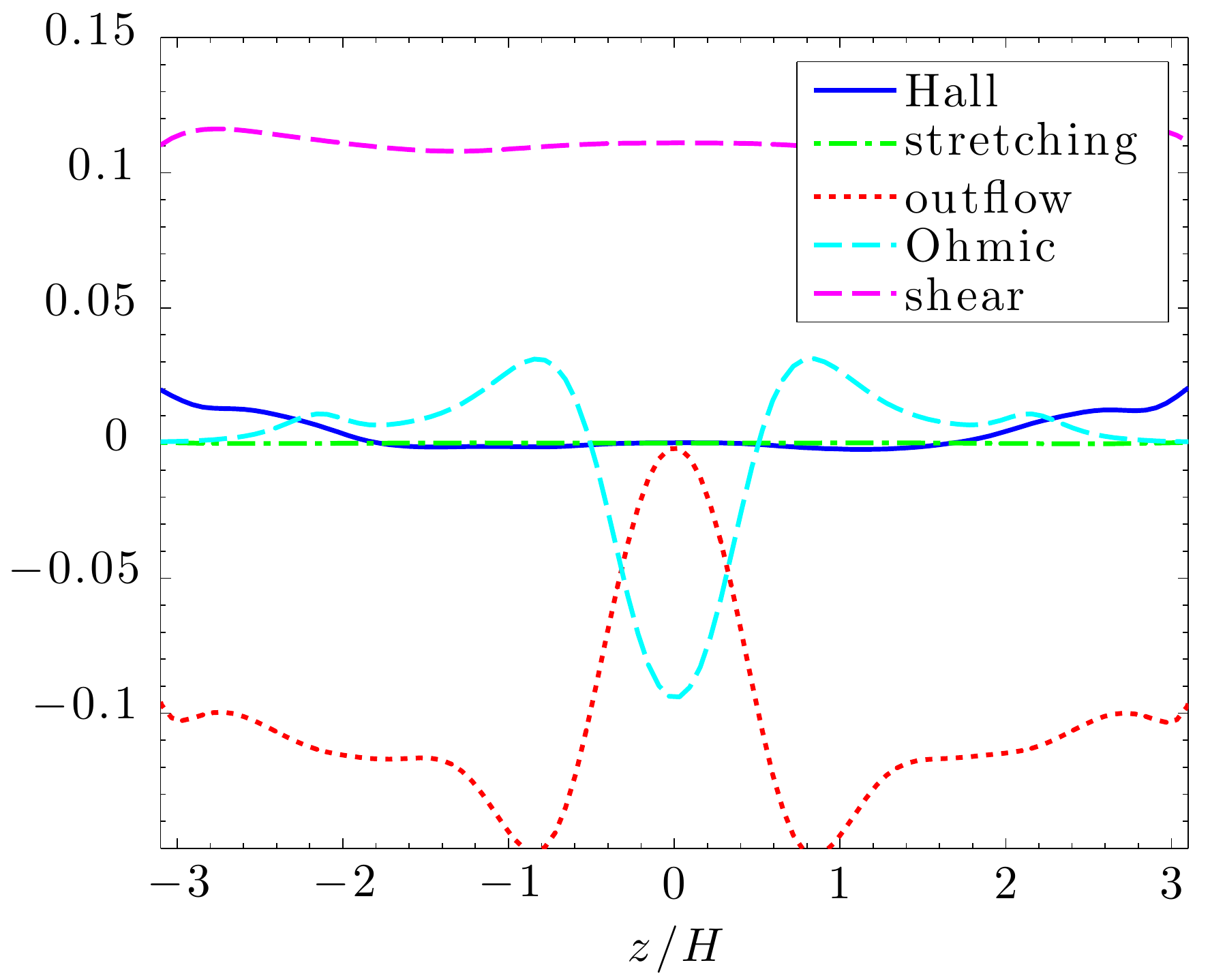}
\caption{Contributions in run 1-OH-5 to the right-hand side of the horizontally averaged induction equation deduced from Equations (\ref{eq:xind}; top) and 
(\ref{eq:yind}; bottom).}
\label{fig:induction}
\end{figure}

The presence of a very strong mean azimuthal magnetic field in the disc midplane of our Hall-MHD simulations is surprising. 
To understand its origin, we decompose the horizontal components of the horizontally averaged induction equation (eq.~\ref{eq:induct}) 
into several pieces, each of which is identified with a physical process:
\begin{subequations}\label{eq:xyind}
\begin{eqnarray}
\pD{t}{\langle B_x \rangle} &=&\underbrace{\pD{z}{\langle B_x v_z \rangle}}_\mathrm{outflow} - \underbrace{\pD{z}{
\langle B_z v_x \rangle}}_\mathrm{stretching}-\underbrace{\pD{z}{}\frac{c(\langle J_xB_z \rangle - \langle J_zB_x \rangle)}{\sqrt{4\pi}en_e}}_
\mathrm{Hall}
\nonumber\\*
\label{eq:xind}
\mbox{} & + &\underbrace{\pD{z}{} \eta \pD{z}{\langle B_x \rangle}}_\mathrm{Ohmic} , \\*
\pD{t}{\langle B_y\rangle}&=&\underbrace{\pD{z}{
\langle B_y v_z \rangle}}_\mathrm{outflow} - \underbrace{\pD{z}{\langle B_z v_y \rangle}}_\mathrm{stretching} -\underbrace{\pD{z}{} \frac{c(\langle J_yB_z \rangle - \langle J_zB_y \rangle)}{\sqrt{4\pi}en_e}}_\mathrm{Hall}
\nonumber \\*
\label{eq:yind}
 \mbox{} & +&\underbrace{\pD{z}{} \eta \pD{z}{\langle B_y \rangle}}_\mathrm{Ohmic} - \underbrace{q\Omega \langle B_x\rangle}_\mathrm{shear}.
\end{eqnarray}
\end{subequations}
We have identified five types of terms:
\begin{description}
\item[\emph{outflow}] describes the vertical transport of horizontal flux tubes. It measures the rate at which the 
outflow evacuates horizontal magnetic-field lines from the disc midplane;
\item[\emph{stretching}] quantifies the horizontal bending of vertical field lines due to horizontal motions;
\item[\emph{shear}] measures the creation of azimuthal field from radial field by the mean Keplerian shear;
\item[\emph{Hall}] denotes the contributions from the Hall effect; and
\item[\emph{Ohmic}] denotes the contributions from Ohmic dissipation.
\end{description}
The saturated state of run 1-OH-5 is shown in Fig.~\ref{fig:induction}, with each of these terms identified.\footnote{The Hall term is artificially large in Fig.~\ref{fig:induction}-top. 
This is because of the capping procedure used in the code (\S\ref{sec:ionisation}): highly magnetised cells see their Hall diffusivities artificially decreased, an effect we can only approximate 
when post-processing horizontally averaged quantities.} We focus on the inner disc region $|z|<3H$ where Hall and Ohmic diffusion dominate the dynamics. 

We first describe the contributions to the right-hand side of the $x$-component of the induction equation (Fig.~\ref{fig:induction}-top). We find that the stretching term is totally negligible and that the 
equilibrium is dominated by a balance between the Hall term, which amplifies the radial field (note that $B_x<0$ in the disc midplane), 
and the outflow term, which carries $B_x$ out of the midplane. This quasi-equilibrium is different than that found in ideal-MHD simulations, in which 
the driving term is due to stretching. On the other hand, the $y$-component of the induction equation (Fig.~\ref{fig:induction}-
bottom) is a resistive MHD quasi-equilibrium, in which the shear production of azimuthal field from radial field is balanced by the outflow term in the atmosphere 
and the Ohmic term in the disc midplane. The Hall terms are completely negligible here.

This analysis allows us to present a self-consistent picture of run 1-OH-5. We first note that
\begin{align}
\langle J_xB_z \rangle - \langle J_zB_x \rangle &=-\partial_z \langle B_yB_z\rangle ,\nonumber\\
\langle J_yB_z \rangle - \langle J_zB_y \rangle &=+\partial_z \langle B_xB_z\rangle .\nonumber
\end{align}
Moreover, since the fluctuations $\delta B_z$ are small compared to the mean vertical field $B_{z0}$, we have $\partial_z \langle B_yB_z \rangle \simeq B_{z0} \partial_z \langle B_y\rangle$ and $\partial_z \langle B_xB_z\rangle\simeq B_{z0}\partial_z \langle B_x\rangle$. Retaining only the driving terms in Equation (\ref{eq:xyind}) and neglecting the outflow, we find
\begin{subequations}\label{eq:red_indxy}
\begin{align}
\label{eq:red_indx}\pD{t}{\langle B_x\rangle }&\simeq B_{z0}\pD{z}{} \frac{c}{\sqrt{4\pi} e n_{\rm e}}\pD{z}{} \langle B_y\rangle, \\
\label{eq:red_indy}\pD{t}{\langle B_y\rangle }&\simeq - \left( B_{z0}\pD{z}{} \frac{c}{\sqrt{4\pi} e n_{\rm e}} \pD{z}{} + q\Omega\right) \langle B_x\rangle .
\end{align}
\end{subequations}
These reduced equations for $\langle B_x \rangle$ and $\langle B_y \rangle$ describe a non-local version of the ``Hall-shear'' instability \citep[cf.][eq.~46]{K08}: Keplerian shear generates an azimuthal magnetic-field component from a radial one, and the Hall effect conservatively reorients this azimuthal field back into the radial direction. When $q\Omega B_{z0} > 0$, this feedback loop can lead to growth. Note that the Hall-shear instability is not a version of the MRI and does not rely on the Coriolis force (though it can be obtained from the Hall-MRI dispersion relation by taking the highly diffusive limit -- see \citealt{RK05} and \citealt{WS12}). It is, however, sensitive to the polarity of the vertical magnetic field. This is discussed more extensively in Section \ref{sec:parameters}.

It is possible to understand the physical content of Equation (\ref{eq:red_indxy}) by observing that $\partial_z \langle B_y \rangle = - \langle J_x \rangle = \langle v_{{\rm d},x} \rangle \sqrt{4\pi} en_{\rm e}$ where $\bm{v}_{\rm d}$ is the drift velocity of the electrons relative to the ions. Hence, the term on the right-hand side of (\ref{eq:red_indx}) is simply the magnetic stretching term due to \emph{electron motion}. In protoplanetary discs the number of free electrons can be very small, so that, for a given current density, the drift speed can be large. As a result, the usual stretching term of ideal MHD caused by the bulk motion of the gas is replaced by a stretching term due to the electron motion. The electron drift produces a radial field from the vertical mean field (eq.~\ref{eq:red_indx}), which is then sheared by the Keplerian flow to produce a azimuthal field (eq.~\ref{eq:red_indy}). It is this azimuthal field that sustains the drift through Amp\`ere's law.

In our simulations, the nonlinear saturated state results from a balance between this Hall-shear instability, an outflow which transports the magnetic energy away from $z\sim H$, and Ohmic diffusion which diffuses the field from the midplane to the base of the outflow (see Fig.~\ref{fig:induction}).

%
%
\subsection{\label{sec:themeat}Nonlinear evolution of stratified discs subject to Ohmic, Hall, and ambipolar diffusion}
\subsubsection{Fiducial runs}

%
%
\begin{figure*}
\centering
\includegraphics[width=0.9\linewidth]{figures/T14_Ma.png}
\includegraphics[width=0.9\linewidth]{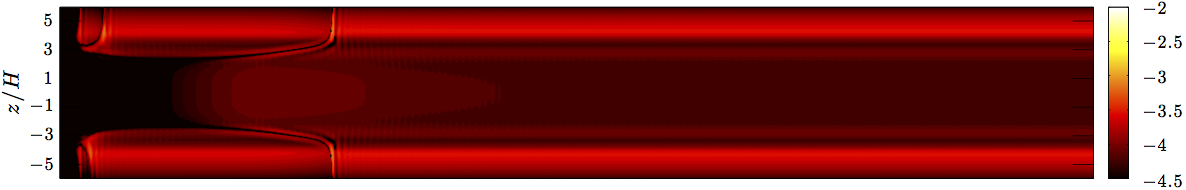}
\includegraphics[width=0.9\linewidth]{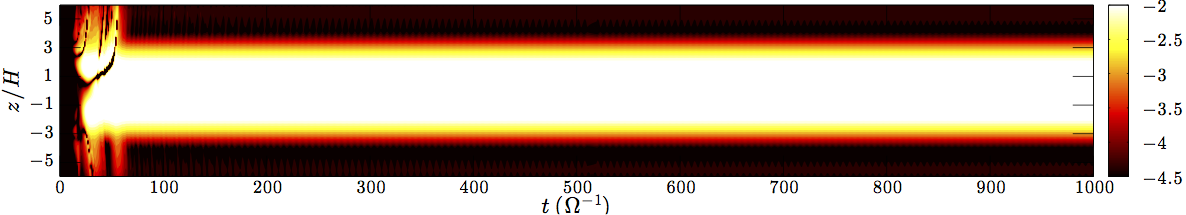}
\caption{Space-time evolution of the logarithm of the horizontally-averaged magnetic stress, $\log\langle M_{xy}\rangle$, in the Ohmic (1-O-5; top), Ohmic-ambipolar (1-OA-5; middle), and Ohmic-ambipolar-Hall (1-OHA-5; bottom) runs.}
\label{fig:ma_OHA}
\end{figure*}

With an understanding of how a strong Hall effect impacts resistive protoplanetary discs, we now turn to simulations also including 
ambipolar diffusion. A comparison of the magnetic stresses 
measured in the Ohmic (1-O-5), Ohmic-ambipolar (1-OA-5), and Ohmic-ambipolar-Hall (1-OAH-5) runs at $R_0 = 1~{\rm au}$ is 
presented in Fig.~\ref{fig:ma_OHA}. The first two simulations are quantitatively similar to runs 
O-b5 and OA-b5 of \cite{BS13}, and we recover their main statistical properties. However, the inclusion of Hall diffusion 
dramatically changes the picture. Similar to the behaviour discussed in Section \ref{sec:nlHall}, we find a large vertical band of strong 
Maxwell stress in the disc midplane, which substantially increases the transport by almost two orders of magnitude.

\begin{figure}
\centering
\includegraphics[width=0.9\linewidth]{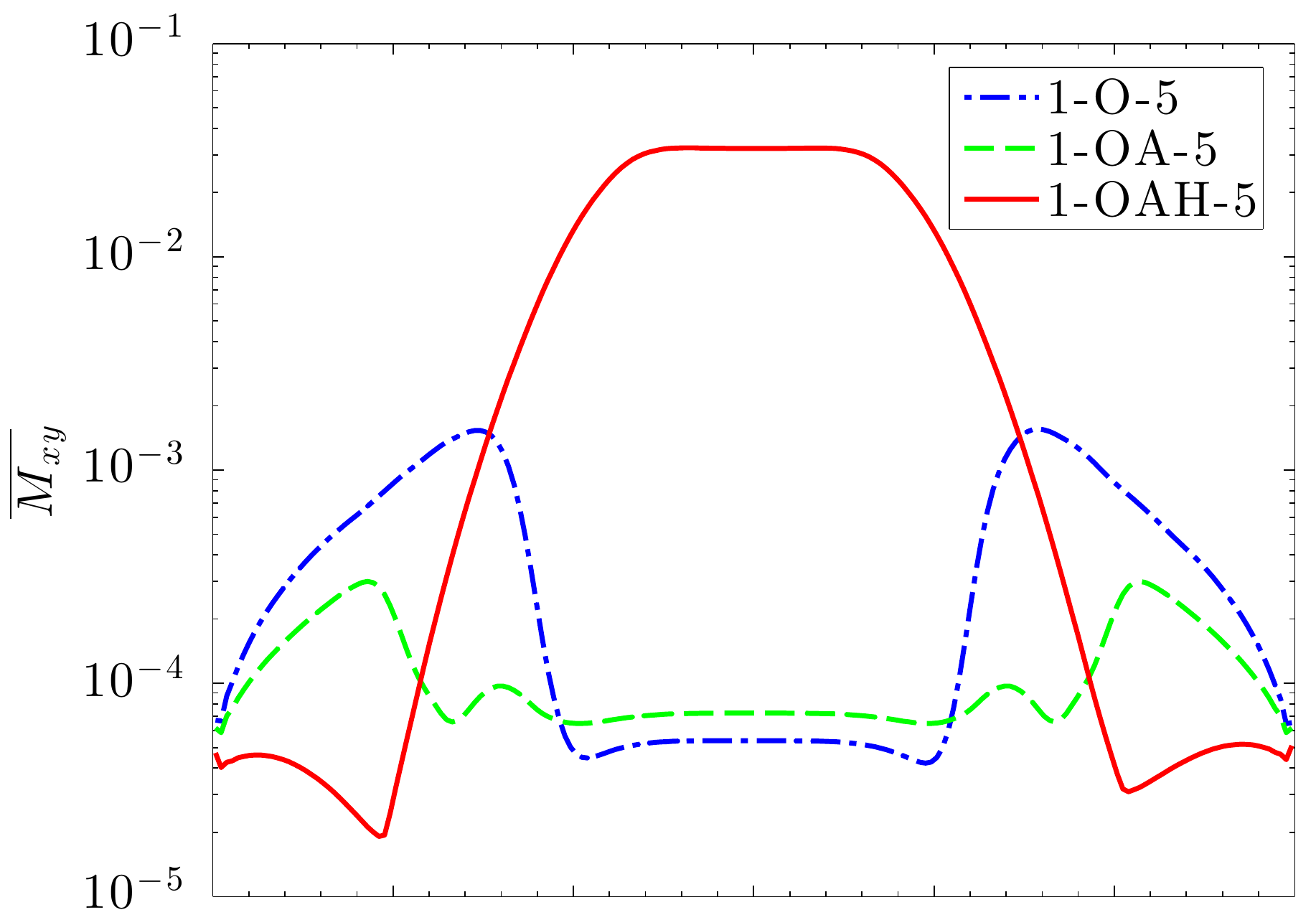}
\includegraphics[width=0.9\linewidth]{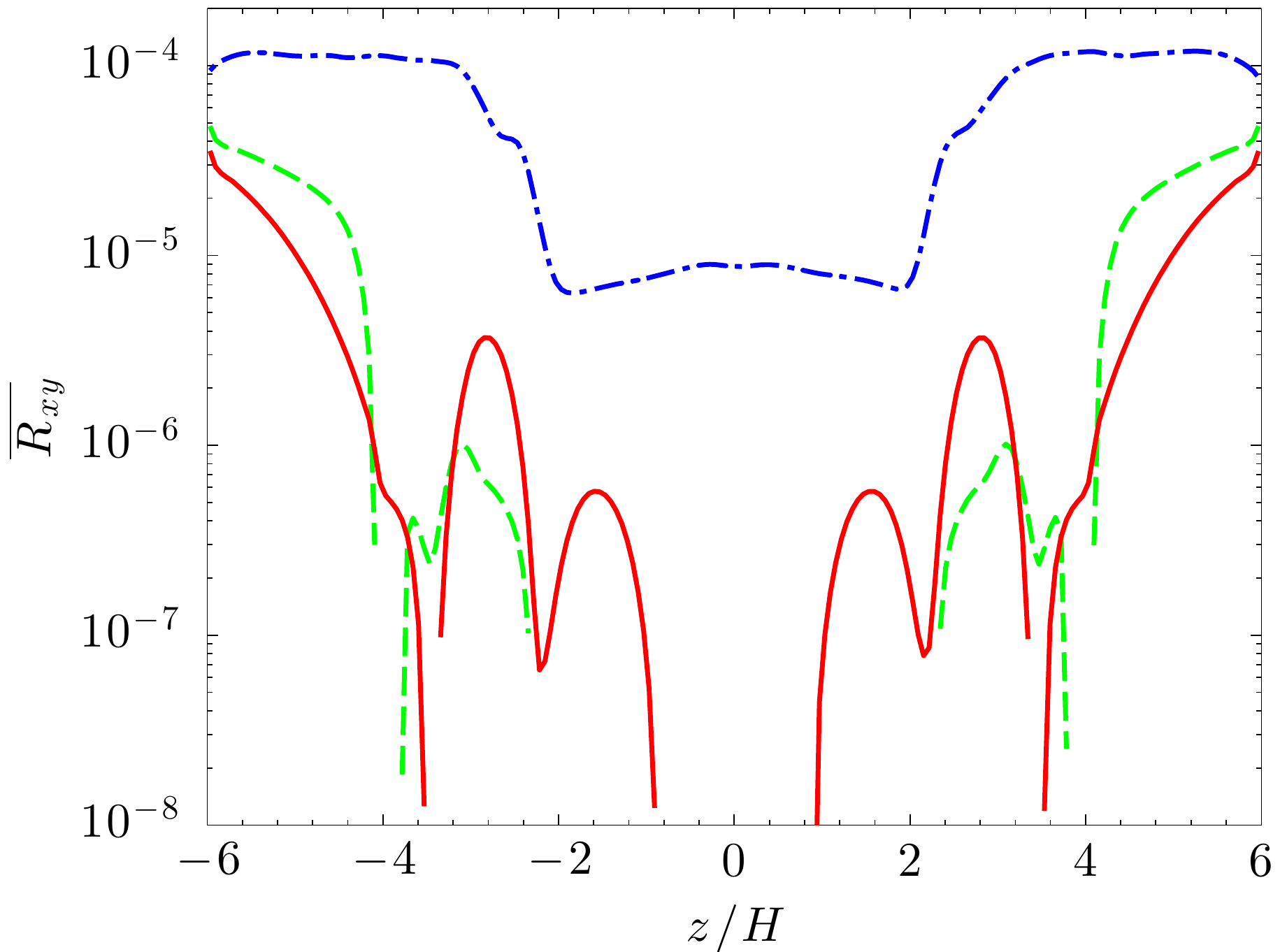}
\caption{Horizontally averaged Maxwell (top) and Reynolds (bottom) stress versus height $z$, averaged over $700\Omega^{-1}$, from runs 1-O-5 (dash-dotted), 1-OA-5 (dashed), and 1-OAH-5 (solid).}
\label{fig:stress_avg}
\end{figure}

The resulting stress profiles averaged over the entire simulation from $t=300\Omega^{-1}$ are presented in Fig.~\ref{fig:stress_avg}. 
The addition of ambipolar diffusion tends to weaken the activity in the surface 
layers \citep[cf.][]{BS13}. The addition of the Hall effect does not alter this picture, and we find that the surface 
activity is even more reduced in the OAH run. However, magnetic activity is greatly enhanced in the midplane, 
where stresses as high as $M_{xy}=3\times10^{-2}$ are found. The Reynolds stress is not significantly 
affected by the Hall effect. This implies that the flow is not ``turbulent'' in the usual sense. 

\begin{figure}
\centering
\includegraphics[width=0.9\linewidth]{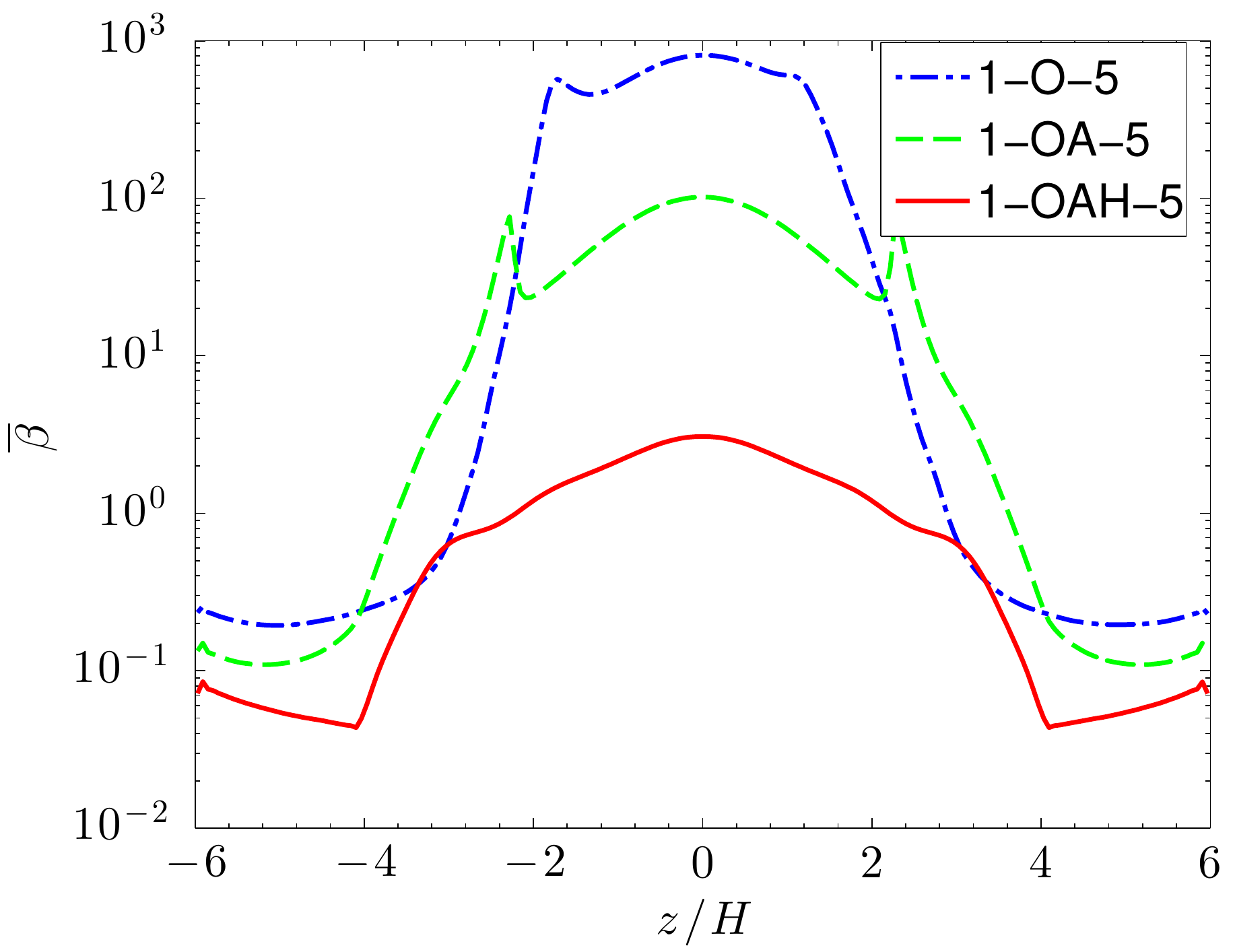}
\caption{Horizontally averaged turbulent plasma $\beta$ versus height $z$, averaged over $700\Omega^{-1}$, from runs 1-O-5 (dash-dotted), 1-OA-5 (dashed), and 1-OAH-5 (solid).}
\label{fig:beta_avg}
\end{figure}

The presence of a strong Hall effect in the midplane also affects the vertical structure of the disc. We find a turbulent
plasma beta $\overline{\beta}\sim 1$ in the disc midplane in run 1-OAH-5 (Fig.~\ref{fig:beta_avg}). The hydrostatic equilibrium
is significantly affected by the magnetic structure and the disc scale height increases, as in Section \ref{sec:nlHall}. Outflows are also affected in this new quasi-equilibrium state, producing mass-loss rates a factor of 20 larger than those obtained in the Ohmic-ambipolar simulations.

%
%

%
%

%
%
\subsubsection{Parameter study}
\label{sec:parameters}
\paragraph{Field polarity:} 

It is well known that the linear Hall-MRI is sensitive to the field polarity $\mathcal{P}=\mathrm{sign}(\beta_0)$ \citep{W99,BT01}. 
To test whether or not this result carries over to the nonlinear regime, we have performed a simulation identical to run 1-OHA-5 
but with a reversed mean field: $\mathcal{P}=-1$ (run 1-OHA-mB). In this case, we find that the level of 
transport, the surface stress, and the mass-loss rate is smaller than in the run without the Hall effect (run 1-OA-5). Therefore, when $\mathcal{P}=-1$, the Hall effect 
\emph{weakens} the efficiency of the MRI and the strength of the outflow. This sensitivity to the field polarity was already 
pointed out in Section \ref{sec:Hall-induction} as a natural consequence of the fact that the radial drift of electrons drives 
the nonlinearly saturated state. A similar trend was found by \cite{SS02} in unstratified shearing boxes, though the 
Hall effect was much weaker in their case and the result less extreme.

\paragraph{Zero vertical-net-flux configuration:}

The zero vertical-net-flux configuration has been largely explored in the past, both with unstratified \citep{HGB96,FP07} and stratified simulations \citep{BNST95,SHGB96,FS03,SBA12,SB13a}. 
We have only explored one such situation including all three non-ideal effects (run 1-OHA-znf). This 
simulation is identical to run 1-OHA-5 except that the initial magnetic-field configuration is given by $B_z=B_{z0}\sin(2\pi x/L_x)$, where $B_{z0}$ is the 
initial field strength from run 1-OHA-5. This initial condition rapidly collapses into a quiet state with virtually no transport 
and no outflow (see Table \ref{tab:runs}).

\paragraph{Mean field strength:}

\begin{figure}
\centering
\includegraphics[width=0.9\linewidth]{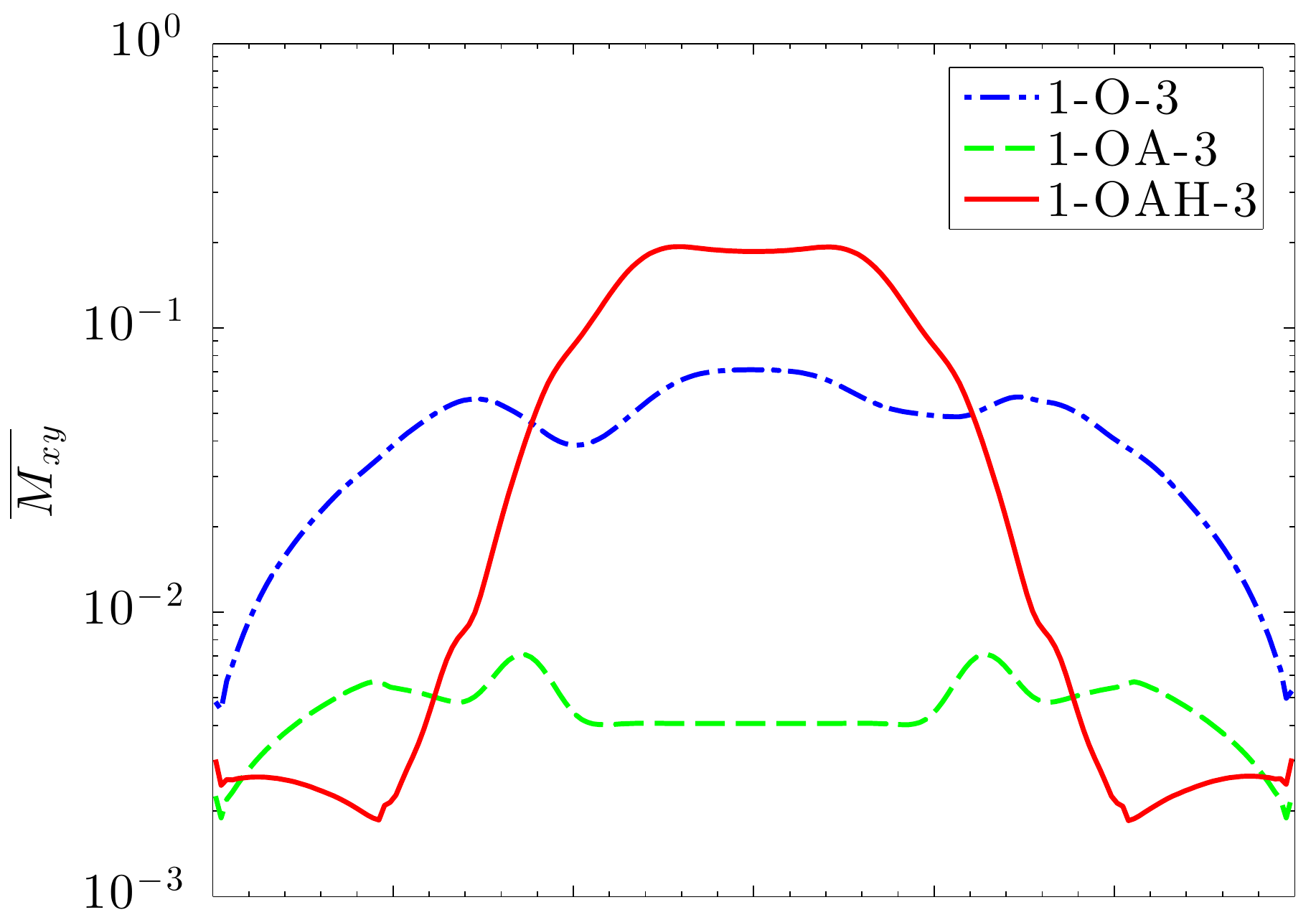}
\includegraphics[width=0.9\linewidth]{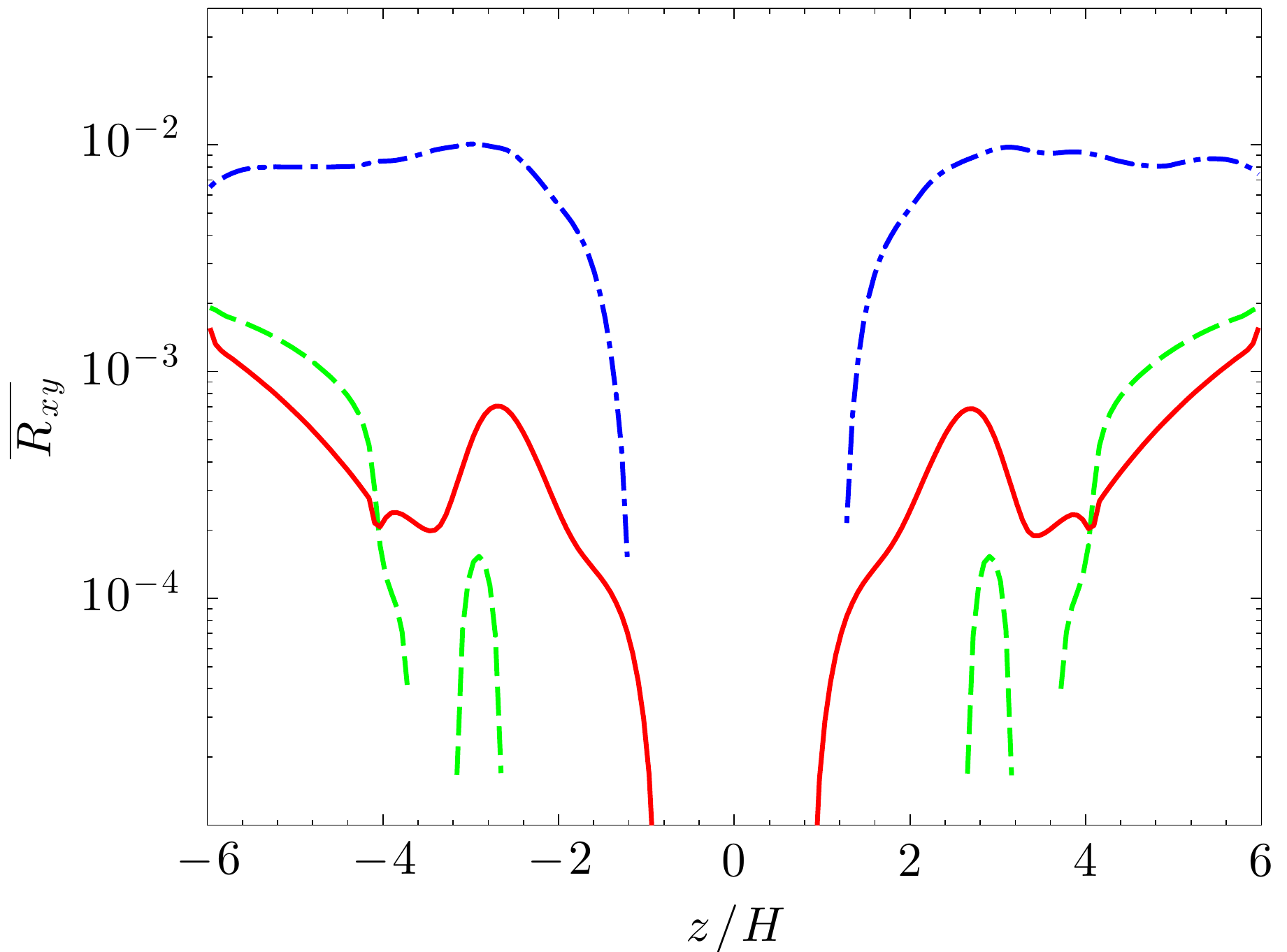}
\caption{Horizontally and temporally averaged Maxwell (top) and Reynolds (bottom) stress versus height $z$ for runs 1-O-3 (dash-dotted), 1-OA-3 (dashed), and 1-OAH-3 (solid).}
\label{fig:stress_avg_3}
\end{figure}

The runs 1-xxx-5 have an initially weak mean vertical magnetic field in the disc midplane. To test how this initial condition might influence 
the subsequent evolution, we have run a set of simulations with $
\beta_0=10^3$ (runs 1-O-3, 1-OA-3, 1-OAH-3). The resulting profiles are presented in Fig.~\ref{fig:stress_avg_3}, 
which may be directly compared to the weak field case (Fig.~\ref{fig:stress_avg}). An initially stronger mean field increases 
significantly the transport in the active surface layers of the Ohmic run, as anticipated using results from ideal-MHD simulations \citep[cf.][]{HGB95}. 
A strong Maxwell stress also appears in the disc midplane, most likely due to horizontal field lines diffusing down  
from the active layers \citep{TS08}. Taken as a whole, these three simulations exhibit  
Maxwell stresses in the midplane that are roughly one order of magnitude larger than those found in run 1-OHA-5; this scaling is qualitatively similar 
to the $\langle M_{xy} \rangle \propto \beta_0^{-1/2}$ scaling found in ideal-MHD simulations \citep{HGB95}. Note, however, that $\alpha$ (the vertically integrated stress) 
does not increase that steeply (see Table \ref{tab:runs}); the stress is less vertically distributed for $\beta_0=10^3$. Fitting our results with a simple power law, we obtain $\alpha\propto \beta_0^{-0.4}$.

The outflow is also affected by the mean field strength. We recover approximately the scalings $\dot{M}_{\rm outflow}\propto \beta_0^{-0.54}$ and 
$T_\mathrm{yz}^S\propto\beta_0^{-0.7}$ found by \cite{BS13} in runs including only Ohmic and ambipolar diffusion. When Hall effect is included, 
the scaling for $T_\mathrm{yz}^S$ still holds but the scaling for $\dot{M}_{\rm outflow}$ is shallower with $\dot{M}_{\rm outflow}\propto\beta_0^{-0.3}$.

%
%

\paragraph{Distance from the central protostellar object:}

\begin{figure}
\centering
\includegraphics[width=0.9\linewidth]{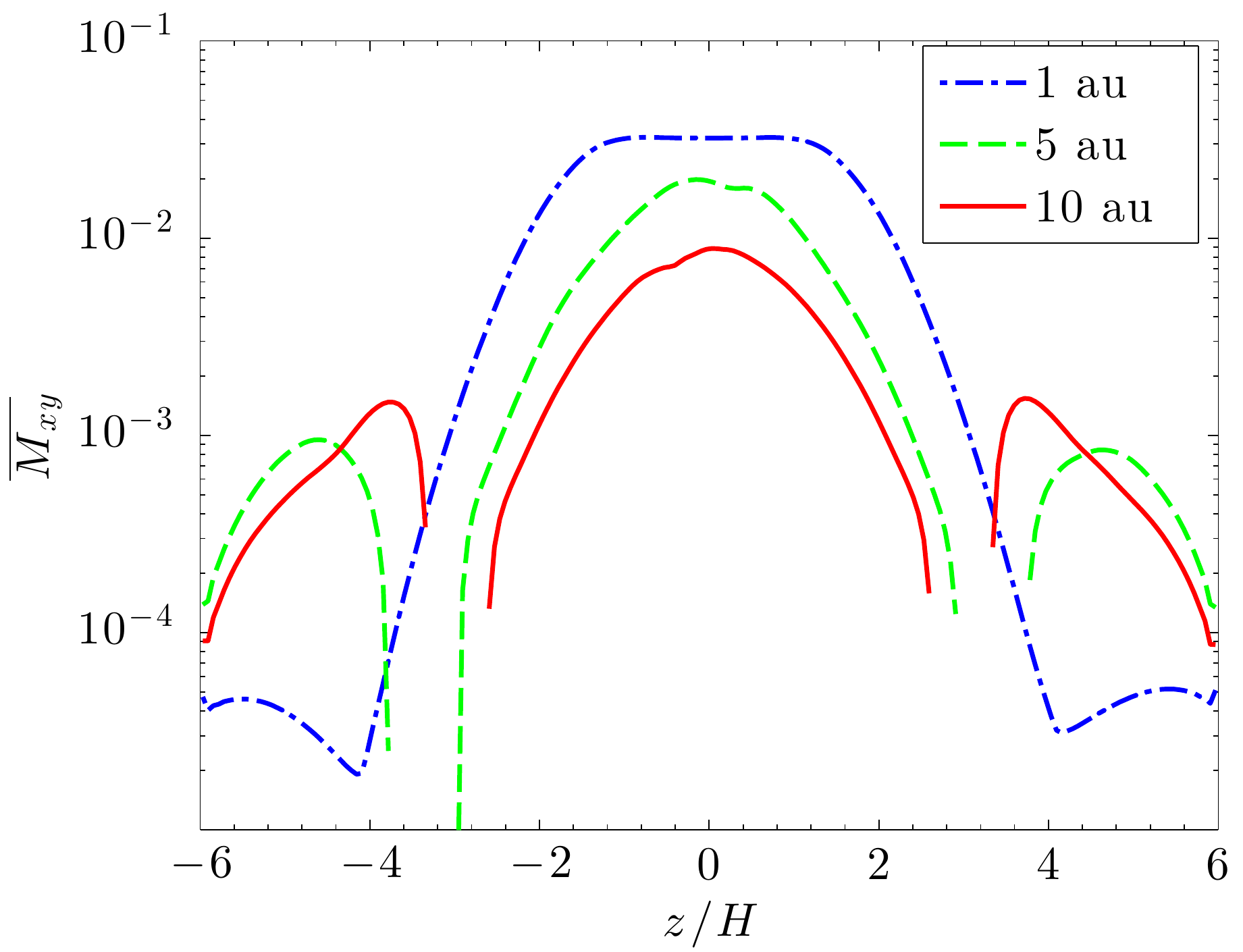}
\caption{Horizontally averaged Maxwell stress versus height $z$, averaged over $700\Omega^{-1}$, from runs 1-OAH-5 
(dash-dotted), 5-OAH-5 (dashed), and 10-OAH-5 (plain line).}
\label{fig:stress_AU}
\end{figure}

The ionisation profile depends strongly upon the radial location in the disc, $R_0$ (\S \ref{sec:ionisation}). At larger 
distances the gas density is lower, and not only do cosmic rays penetrate deeper into the disc interior, but also the recombination time is longer. Both effects lead to 
an increase in the ionisation fraction. For this reason, the saturation properties of the MRI vary with radius. We have performed three simulations with $\beta_0=10^5$, varying $R_0$ from 1 au to 10 au. The resulting Maxwell stress profiles are presented in Fig.~\ref{fig:stress_AU}.\footnote{The Reynolds stress is negligible in these runs.} 

We find that the Maxwell stress in the disc midplane decreases as one moves to larger radii, since $\rm{Ha}$ increases and approaches $\rm{Am}$. 
Note that the outflow surface stress and the mass-loss rate decrease in a similar fashion (see Table \ref{tab:runs}). By contrast, 
the Maxwell stress tends to increase in the surface layers, contributing roughly $1/5$ of the vertically integrated stress at 10 au.
We find that this component of the stress at 5 and 10 au is due to turbulence in the surface layers, in a way qualitatively similar to that found in 
run 1-OH-5. Overall, this trend indicates that, as we move outward, we approach the point at which the stress contribution due to the surface layers dominates 
that of the vertically integrated disc. This corresponds to the classic layered accretion picture, a limit that corresponds to ``region III'' described by \cite{B13} 
and studied extensively by \cite{SB13,SB13a} in situations where ambipolar diffusion suppresses the MRI in the disc midplane.

%
%

\paragraph{Symmetries:}

\begin{figure*}
\centering
\includegraphics[width=0.9\linewidth]{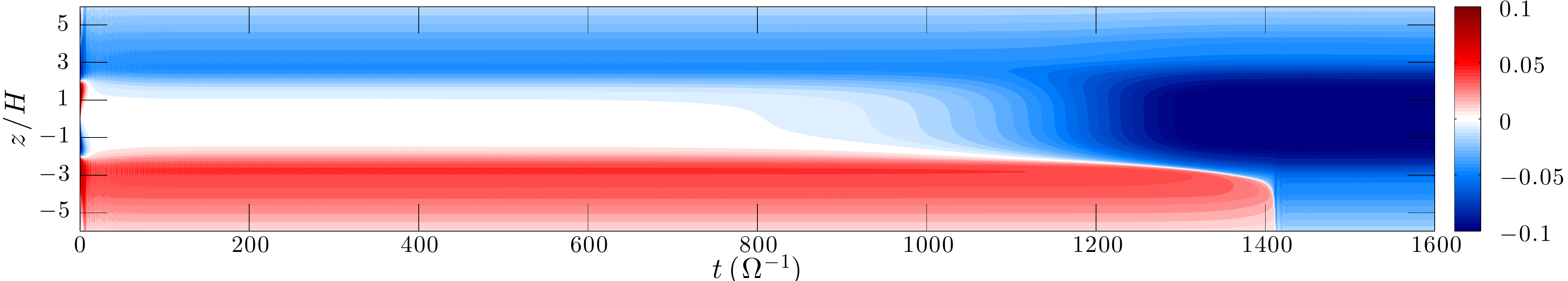}
\caption{\label{fig:by_sim}Space-time evolution of the horizontally averaged azimuthal magnetic field, $\langle B_{y}\rangle$, as a function of time in run 1-OA-5-e. The run starts 
with odd symmetry for the azimuthal field. At $t\simeq 1000\Omega^{-1}$ a current layer forms and is ejected at $t=1410\Omega^{-1}$, leaving a azimuthal field with even symmetry.}
\end{figure*}

The saturated states of all the OAH runs exhibit the same $\sigma=1$ symmetry (see eqns~\ref{eq:symmetries}). 
However, this symmetry implies that the outflow points inwards on one side of the disc and outwards on the other side. 
A more physical outflow solution has $\sigma=-1$. It has been shown by \cite{BS13} that ambipolar-diffusion--dominated 
simulations can exhibit solutions having outflows with the proper symmetry at $|z|\gtrsim 
3.5H$ if particular initial conditions are chosen. These solutions exhibit a strong \emph{off-midplane} current 
layer, which was found to be stable by \cite{BS13}. We have tried to find similar solutions including the Hall effect but have 
failed: we have only found $\sigma=1$ solutions when Hall diffusion is included. However, we are able 
to find long-lived $\sigma=-1$ solutions when only Ohmic and ambipolar diffusion are included (see run 1-OA-5-e in 
Fig.~\ref{fig:by_sim}). These solutions start to develop a strong off-midplane current layer (around $t
\sim1000\Omega^{-1}$), which is very similar to the one presented by \cite{BS13}. However, we find that this layer is eventually ejected 
on longer timescales and the system ultimately relaxes into a $\sigma=1$ symmetry. This casts doubt upon the stability and even the applicability  
of these outflow solutions in actual protoplanetary discs.

%
%

%
%
\subsubsection{Magnetic self-organisation in Hall-dominated magnetorotational turbulence}

KL13 have shown that the saturated state favoured by Hall-dominated magnetorotational turbulence 
in unstratified shearing boxes with net vertical magnetic flux is characterised by a strong axisymmetric (``zonal'') magnetic field and a vanishingly low level of turbulent 
transport. Remarkably, none of the simulations presented in this paper exhibit this behaviour. This disparity is not caused by the different numerical algorithms employed. 
Indeed, we were able to reproduce their results with our fiducial 16 points per $H$ (see \S\ref{sec:hall-unstrat}). Instead, the difference is due to the strong azimuthal field that is naturally produced in our stratified simulations. In unstratified simulations, the net magnetic flux is conserved and, if there is initially no net azimuthal flux, none will be generated ($\langle B_y \rangle = 0$). 
By contrast, the outflow boundary conditions imposed at the top and bottom of the stratified shearing box allows a net azimuthal field 
to develop: azimuthal flux of opposite polarity is ejected during the generation of the outflow. As a result, stratified simulations with the Hall effect can produce a 
(\emph{very} large) mean azimuthal field relative to the mean vertical field, typically with $\langle B_y\rangle\sim 200\langle B_z\rangle$. As pointed out by KL13, 
this magnetic configuration ($\langle B_y \rangle \gg \langle B_z \rangle$) does not saturate via the production of zonal magnetic fields. 

%
%
\section{\label{sec:grains}Influence of dust grains}

Dust grains comprise $\sim$$1\%$ by mass of protostellar cores and, by extension, protoplanetary discs \citep{H81}. In the latter, micron-sized 
grains can dramatically increase the rate of magnetic diffusion, mainly because of their propensity to soak up free charges at high densities. This tends to decouple the gas from the magnetic field \citep{SWH04,W07,B11}. Since these models usually involve complex chemical networks and various grains distribution, it is difficult to obtain a good 
physical intuition for the effect grains can have on the physics of accretion disc turbulence. To clarify the situation, we discuss the qualitative effects of dust 
grains on the chemistry and dynamics of a protoplanetary disc. Our discussion focuses on the disc midplane, where the ionisation rate is very low (typically $\sim$$\mathcal{O}(10^{-14})$).

%
%
\subsection{Diffusivity tensor}
Dust grains preferentially capture free electrons, since electrons have less inertia than the ions. This process occurs quickly, 
so that grains are usually the dominant negative charge carriers when they are well-mixed with the gas. Since grains carry negative charges, 
they also tend to increase the effective recombination rate with ions, acting as a catalyst for recombination. This decreases the total ionisation fraction by one 
or two orders of magnitude, depending upon the grain size. A typical situation is presented in \citet[][fig.~1]{B11}, where we clearly see that the presence of 
$0.1\mu \mathrm{m}$-sized grains decreases the ionisation fraction and makes singly charged grains the dominant charge carriers.

In addition to the modification of the ionisation equilibrium, dust grains also have an impact on the gas dynamics.
From a plasma point of view, the presence of charged grains 
indicates that some charge carriers are much heavier than electrons and ions. This suggests that the average mobility of 
charge carriers and their coupling time with neutral $\mathrm{H}_2$ are drastically reduced. The diffusivities associated to the three non-ideal MHD effects ($\eta_{\rm O}$, $\eta_{\rm H}$, $\eta_{\rm A}$) are therefore significantly altered.

In order to compute the diffusivities in such a plasma, one should follow the derivation of the \emph{complete} diffusivity tensor of 
\cite{W07}, which takes into account all charged species. This calculation is beyond the scope of this paper, but we can rely on the work of \cite{SW08} to investigate the impact of dust grains on protoplanetary disc midplanes. Consider their fig.~1, which presents diffusivity profiles for various grain contents. Without grains, we observe that, at 5 and 10 au, the dominant non-ideal effect in the midplane is Hall, which matches our own diffusivity profiles (see \S\ref{sec:ionisation}). When 1$\mum$ grains are introduced, the respective ratios of Ohmic, Hall, and ambipolar diffusion are largely unaffected in the midplane, although each of these diffusivities are increased by roughly three orders of magnitude. The most dramatic modification comes from $0.1\mum$ grains: at 10 au, those authors found that ambipolar diffusion dominates in the disc midplane, followed by the Hall effect at about a factor of 10 smaller. At 5 au, all three diffusivities are comparable in the midplane, although their absolute values are increased by five to six orders of magnitude.

%
%
\subsection{Vertical distribution}
\label{sec:vert_distribution}
Despite the presence of a strong magnetic torque and the consequent large rates of angular-momentum transport observed in our Hall-dominated 
simulations, the flow appears to be predominantly laminar. One might then expect any population of large dust grains to slowly settle into the disc 
midplane and thereby affect the diffusivity tensor. However, the driven outflows may lift up these grains and, in doing so, mimic the role 
traditionally ascribed to turbulent stirring. It is therefore natural to ask whether or not grains are expected to sediment under the conditions 
found in our simulations.

To this end, let us consider the vertical equilibrium for dust grains obtained by balancing vertical gravity with gas drag:
\begin{equation}
\nonumber 0 = - m_{\rm g} g_z + \langle \sigma v \rangle_{\rm g} \, \rho v_z ,
\end{equation}
where $g_z$ is the vertical component of gravity, $v_z$ is the vertical gas velocity, and $\langle \sigma v \rangle_{\rm g}$ is the rate of momentum exchange of grains with neutral gas molecules. We approximate $\langle \sigma v \rangle_{\rm g} \sim a^2 c_{\rm s}$ and $m_{\rm g} \sim a^3 \rho_{\rm S}$, where $\rho_{\rm S}$ is the grain material density and $a$ is the grain radius. Using $g_z = \Omega^2 z$, we find
\begin{equation}
\nonumber z_{\rm max} \sim \frac{\rho}{\rho_{\rm S}} \frac{c_{\rm s} v_z}{a \Omega^2}
\end{equation}
to be the equilibrium height reached by the grains when they are dragged by the outflow. The continuity equation implies that $\rho v_z$ is conserved along $z$, defining the mass-outflow rate. Assuming $\rho_{\rm S} = 1~{\rm g~cm}^{-3}$ and using Equation \ref{eqn:mdot} to quantify the mass-loss rate, we obtain
\begin{equation}
\nonumber \frac{z_{\rm max}}{H} \sim 10^7 \left( \frac{\Sigma}{10^3~{\rm g~cm}^{-2}} \right) \left( \frac{a}{1~\mu{\rm m}} \right)^{-1}~\dot{M}_{\rm outflow} .
\end{equation}
This estimate suggests that, with typical outflow mass-loss rates found in our simulations (i.e.~$\dot{M}_{\rm outflow} \gtrsim 10^{-4}$), dust grains with sizes up to 1 mm (30 $\mu$m) can be lifted up at 1 au (10 au).

%
%
\subsection{Impact on this work}

Overall, it appears that $1\mum$ grains should not have too much of an impact on our results at 5 and 10 au: the Hall effect still dominates in the midplane, and the increase in the diffusivities by three orders of magnitude leads to Elsasser numbers close to the midplane values we adopted at $R_0 = 1~{\rm au}$. Hence, the role played by the Hall effect is likely to be qualitatively comparable to what we presented in Section \ref{sec:themeat}, though quantitatively different. To test this conjecture, we performed two simulations at 10 au in which the diffusivities are artificially increased by a factor of 1000 throughout the vertical extent of the disc (runs 10-OA-gr and 10-OHA-gr). Even in this worst-case scenario, the Hall effect still enhances significantly the magnetic stress in the disc (see $\alpha$ values in Tab.~\ref{tab:runs}).

 For $0.1\mum$ grains, the impact is less obvious. The Hall effect can still be a major player at 5 au, but the dramatic increase of the diffusivities brings us into new territory, for which we have no numerical simulations to guide our intuition. This regime is quite difficult to study numerically, since the diffusivities are much larger than the cap values we used in this work (see \S\ref{sec:ionisation}). Increasing the cap values to the predicted diffusivities would lead to extremely small time steps, making explicit numerical simulations such as ours impractical at this time.

%
%
\section{\label{sec:conclusions}Summary}

In this paper we have explored the linear and nonlinear behaviour of poorly ionised, vertically stratified protoplanetary discs. For 
the first time, all three relevant non-ideal MHD effects -- Ohmic dissipation, ambipolar diffusion, and the Hall effect -- are self-consistently included. 
To accomplish this feat, we have implemented an original formulation of ambipolar diffusion and the Hall effect in the finite-volume code \textsc{Pluto}.

Our results demonstrate that none of these effects can be safely neglected at distances $\sim$1--10 au from the central protostellar object. En route to this conclusion, 
we have confirmed previous work on the effects of Ohmic dissipation and ambipolar diffusion on magnetorotational turbulence in stratified discs: 
Ohmic diffusion quenches the MRI in the disc midplane and ambipolar diffusion suppresses turbulence in the surface layers, making the traditional viscous disc model 
insufficient to explain the observed accretion rates ($\alpha \approx 6 \times 10^{-4}$ at 1 au). Our principal finding is that the Hall effect alters this picture dramatically. 
A strong azimuthal magnetic field is produced in the disc midplane, which generates a vertical magnetic pressure gradient strong enough to substantially increase 
the vertical scale height of the disc. This azimuthal field correlates with a weak radial field to produce very efficient angular-momentum transport, with $
\alpha \sim 10^{-2}$. These values are compatible with accretion rates $\dot{M} \lesssim 10^{-7}~{\rm M}_\odot~{\rm yr}^{-1}$ (eq.~\ref{eq:mdot}) without a need to invoke external 
stresses generated by disc outflows \citep[as in][]{BS13}.

MRI-driven outflows are also modified by the Hall effect. The mass-loss rates increase by a factor $\lesssim$$20$ over those found in runs with only Ohmic dissipation and ambipolar diffusion. The surface 
stresses also increase, by a factor of $\sim$4 at 1 au. However, we caution that these numbers may be inappropriate for global disc models, since outflows (and in particular the mass-loss rates) are affected by the vertical boundary conditions in the shearing-box approximation \citep{FLLO13,LFO13}. Shearing-box outflows should be treated with caution.

Surprisingly, our stratified simulations do not produce the axisymmetric (``zonal'') magnetic-field structures and the attendant steep reduction in turbulent transport found by KL13 in unstratified simulations with net vertical magnetic flux. This is despite our simulations being in the right regime ($\ellH\gtrsim 0.2 H$). We speculate that this difference is due to the strong azimuthal magnetic field produced in the disc midplane for the stratified case, which ``hides'' the vertical field from the dynamics. This is related to our choice of vertical boundary conditions, which permits the evacuation of magnetic flux tubes from the computational domain and the production of a net azimuthal field (despite none being present initially). This build-up of azimuthal flux is not possible in unstratified shearing boxes. It is suggestive that KL13 did not find the zonal-field route to saturation in unstratified shearing boxes with $\langle B_y \rangle \gg \langle B_z \rangle$. We therefore conjecture that both the global topology of the magnetic field threading the accretion disc and the radial and vertical ionisation profiles may play a role in determining which saturated state is favoured.

It is well-known that the influence of the Hall effect on the linear stability of the disc depends upon the field polarity $\bm{\Omega\cdot}  \bm{B}$ \citep{W99,BT01}, a result we have found to 
hold true in the nonlinear regime as well \citep[see also][]{SS02}. Configurations with $\bm{\Omega\cdot} \bm{B}  > 0$ show enhanced transport, while configurations with $\bm{\Omega\cdot} \bm{B} < 0$ show  
reduced transport (viz.~$\alpha\sim10^{-4}$ and no significant outflow). This suggests that one could have disconnected regions of the disc with different field polarities, which are either actively accreting or relatively quiet and evolve according to the global topology and the long-term evolution of the large-scale magnetic field.

The relatively large magnetic stresses generated by the Hall effect are not associated with turbulent fluctuations. Rather, they are associated with large-scale amplification of an 
azimuthal magnetic field via the shearing of a Hall-induced radial field \citep[cf.][]{K08}. Indeed, the Reynolds stresses are always very small in the midplane (with 
$\langle R_{xy} \rangle \lesssim 10^{-6}$). This effect is distinct from the process found by \citet{TS08}, in which the magnetic field Ohmically diffuses from the upper active layers 
down into the midplane, where it exerts a large-scale stress. One consequence of our result is that Hall-dominated MRI ``turbulence'' may be unable to stir up dust grains that would 
otherwise slowly settle into the disc midplane under the action of vertical gravity. This could be a problem, since observations have implied that sub-$\mu$m grains are present in the 
upper atmospheres of protoplanetary discs \citep{PP08}. On the other hand, such small grains could have been lifted up by a weak outflow (cf Section \ref{sec:vert_distribution}).

Finally, using \cite{SW08} results, we have shown that $\mum$-sized grains should not have much of an impact on our conclusions: the Hall effect remains the dominant non-ideal term in disc midplanes at 5 and 10 au, with diffusivity coefficients comparable to that at 1 au without grains. 0.1$\mum$-sized grains have a more severe impact on the diffusivities, though Hall remains an important player at 5 au. We have not explored this regime numerically because of the cost of our explicit numerical scheme. We note, however, that there is no reason \emph{a priori} to neglect the Hall effect at these distances. 

While a more sophisticated treatment of the chemistry -- in which dust grains are included and species abundances are time-dependent -- is necessary to make accurate quantitative predictions, the results presented here provide unequivocal evidence that Ohmic dissipation, ambipolar diffusion, and the Hall effect must all be taken into account to obtain a realistic description of angular-momentum--transport in protoplanetary discs.

\begin{acknowledgements}
GL is indebted to Wing-Fai Thi for his many suggestions and clarifications concerning the radiative and chemical processes
relevant to protoplanetary discs. GL also thanks Andrea Mignone and G\'abor T\'oth for fruitful discussions regarding the numerical implementation
of the Hall effect during the Astronum 2013 conference. Support for GL was provided by the European Community via contract PCIG09-GA-2011-294110. 
Support for MWK was provided by NASA through Einstein Postdoctoral Fellowship Award Number PF1-120084, 
issued by the {\it Chandra} X-ray Observatory Center, which is operated by the Smithsonian Astrophysical Observatory for and on behalf of NASA under contract NAS8-03060. 
Support for SF was provided by the European Research Council under the European Union's Seventh Framework Programme (FP7/2007-2013) / ERC Grant agreement n\ts{$\circ$} 258729. 
Most of the the computations presented in this paper were performed using the Froggy platform of the CIMENT infrastructure (https://ciment.ujf-grenoble.fr), which is supported by the Rh\^one-Alpes region (GRANT CPER07\_13 CIRA), the OSUG{@}2020 labex (reference ANR10 LABX56) and the Equip{@}Meso project (reference ANR-10-EQPX-29-01) of the programme Investissements d'Avenir supervised by the Agence Nationale pour la Recherche. This work was granted access to the HPC resources of IDRIS under allocation x2014042231 made by GENCI (Grand Equipement 
National de Calcul Intensif). 
\end{acknowledgements}

%
%
\begin{appendix}

%
%
\section{Numerical implementation of the Hall effect\label{app:hall}}
In this Appendix we detail the modifications made to the \textsc{Pluto} code (v4.0) to include the Hall effect. While these modifications 
are quite general to any conservative finite-volume numerical scheme, the interested reader would benefit by familiarizing themselves with 
both \cite{M07} and the \textsc{Pluto} user manual.

%
%
\subsection{Conservative finite-volume scheme}

The implementation of Hall effect in \textsc{Pluto} has proven to be a difficult task. Our naive attempts to include the Hall 
effect as a source term, in a way analogous to that used for Ohmic diffusion, failed. Since the Hall effect is in essence a 
dispersive term, rather than a true diffusive term, we have instead incorporated the Hall effect into the heart of the conservative 
integration scheme. 

We begin by writing Equations (\ref{eq:cont})--(\ref{eq:induct}) in conservative form:
\begin{equation}\label{eqn:cons}
\frac{\partial \bm{U}}{\partial t}=\bm{\nabla \cdot  F(U)}+\bm{S(U)} ,
\end{equation}
where $\bm{U}$ is a vector of conserved quantities, $\bm{F}$ is the conservative flux function, and $\bm{S}$ is the 
source term function. The equations of isothermal Hall-MHD dictate
\begin{equation}\label{eqn:uandf}
\arraycolsep=1.4pt\def\arraystretch{1.5}
\bm{U}=\left (\begin{array}{c} 
\rho \\
\rho \bm{v}\\
\bm{B}
\end{array}
\right) \quad {\rm and} \quad
\bm{F(U)}=\left( \begin{array}{c}
\rho \bm{v}\\
\rho \bm{v}\bm{v}-\bm{B}\bm{B}-p_{\rm T}\bm{I}\\
\bm{v B}-\bm{B v}-x_{\rm H}(\bm{J B}-\bm{B J})
\end{array}
\right) ,
\end{equation}
where $p_{\rm T} = \rho c^2_{\rm s}-B^2/2$ is the total pressure and $x_{\rm H}=\eta_\mathrm{H}/B$ is 
the control parameter for the Hall effect.

In the above formulation, we have explicitly used $\bm{J}=\bm{\nabla\times B}$ in the 
flux function, which implies that the flux depends upon the conserved variables \emph{and their derivatives}. This has 
dramatic consequences on the mathematical nature of the Riemann problem since it is, in this formulation, ill-defined 
(to wit, if $\bm{B}$ is discontinuous, then $\bm{J}$ is not defined). Another way to see this is to note that the 
linearised flux function, which defines the characteristic speeds of the system, depends upon the wavelength of 
the perturbation (due to the presence of whistler waves). If a discontinuity appears in the flow, one 
obtains infinitely fast characteristic speeds, which are of course unphysical.\footnote{In a realistic plasma, the whistler 
wave speed is limited by finite Larmor radius effects that are excluded in the MHD approximation.} 
Because of this mathematical difficulty, we have not found any sensible way to derive accurate 
Riemann solvers such as the Roe or HLLD solvers. Instead, we follow \cite{T08} and implement an 
HLL solver which approximates the dynamics of Hall-MHD assuming that $\bm{J}$ is an external parameter (i.e.~not 
related to $\bm{B}$). In doing so, we circumvent the difficulties exposed above, at the expense of large numerical 
diffusivities.

The algorithm we designed to integrate Equations (\ref{eqn:cons}) and (\ref{eqn:uandf}) using \textsc{Pluto} is as follows:
\begin{enumerate}
\item Compute the primitive cell-averaged variables $\bm{V}$ from the conserved variables $\bm{U}$.
\item Reconstruct the primitive variables at the cell edges using a ``well-chosen'' second-order total-variation-diminishing 
spatial-reconstruction scheme (see below). This defines a left state ($\bm{V}_{L}$) and a right state ($\bm{V}_{R}$) at each cell face. 
\item Compute the left and right conserved variables $\bm{U}_{R/L}$ from $\bm{V}_{R/L}$.
\item Compute the face-centered $\bm{J}$ from the cell-averaged $\bm{B}$ using finite-difference formulae. Care is taken at this stage 
to apply the proper boundary conditions; shearing-sheet boundary conditions have to be applied explicitly to the currents in order to 
avoid spurious oscillations at radial boundaries due to interpolation errors.
\item Compute the left and right fluxes using the left and right states and the face-centered current: $\bm{F}_{L/R}=
\bm{F}(\bm{U}_{L/R},\bm{J})$.
\item Compute the Godunov flux $\bm{F}^\ast$ using the whistler-modified HLL solver (described below).
\item Evolve in time the conservative variables according to the equations of motion (see \citealt{M07} for details). If 
constrained transport is used \citep{EH88,BS99}, then the Godunov flux is used to compute the edge-centered 
electromotive forces (EMFs), which are then used to evolve the magnetic field.
\end{enumerate}

To compute the Godunov flux $\bm{F}^\ast$, we follow the HLL scheme by using the approximation
\begin{align}
\bm{F}^*&=\bm{F}_L&\quad\mathrm{if\,\,} S_L>0 ;\nonumber\\
\bm{F}^*&=\bm{F}_R&\quad\mathrm{if\,\,} S_R<0 ;\nonumber\\
\bm{F}^*&=\frac{S_RS_L(\bm{U}_R-\bm{U}_L)+S_R\bm{F}_L-S_L\bm{F}_R}{S_R-S_L}&\quad \mathrm{otherwise} ,\nonumber
\end{align}
where $S_L$ is the smallest algebraic signal speed for the left state and $S_R$ is the largest algebraic signal speed 
for the right state. Since we are solving the Hall-MHD equations, $S_{R/L}$ includes both the fast magnetosonic speed and the 
whistler wave speed. Therefore, we choose
\begin{equation}
S=v\pm\mathrm{max}(c_{\rm f},c_{\rm w}),\nonumber
\end{equation}
where $v$ is the flow speed, $c_{\rm f}$ is the fast magnetosonic speed, and $c_{\rm w}$ is the whistler wave speed. As 
whistler waves are dispersive, we choose $c_{\rm w}$ to be equal to the whistler speed at the grid 
scale:
\begin{equation}
c_{\rm w}= \left| \frac{x_{\rm H} B}{2 \Delta x}\right| +\sqrt{ \left(\frac{x_{\rm H} B}{2 \Delta x}\right)^2+\frac{B^2}{\rho}} ,\nonumber
\end{equation}
where $\Delta x$ is the grid spacing in the direction under consideration.

%
%
\subsection{Numerical tests}
%
%
\subsubsection{Linear dispersion relation\label{sec:num_disp}}

\begin{figure}
\centering
\includegraphics[width=0.9\linewidth]{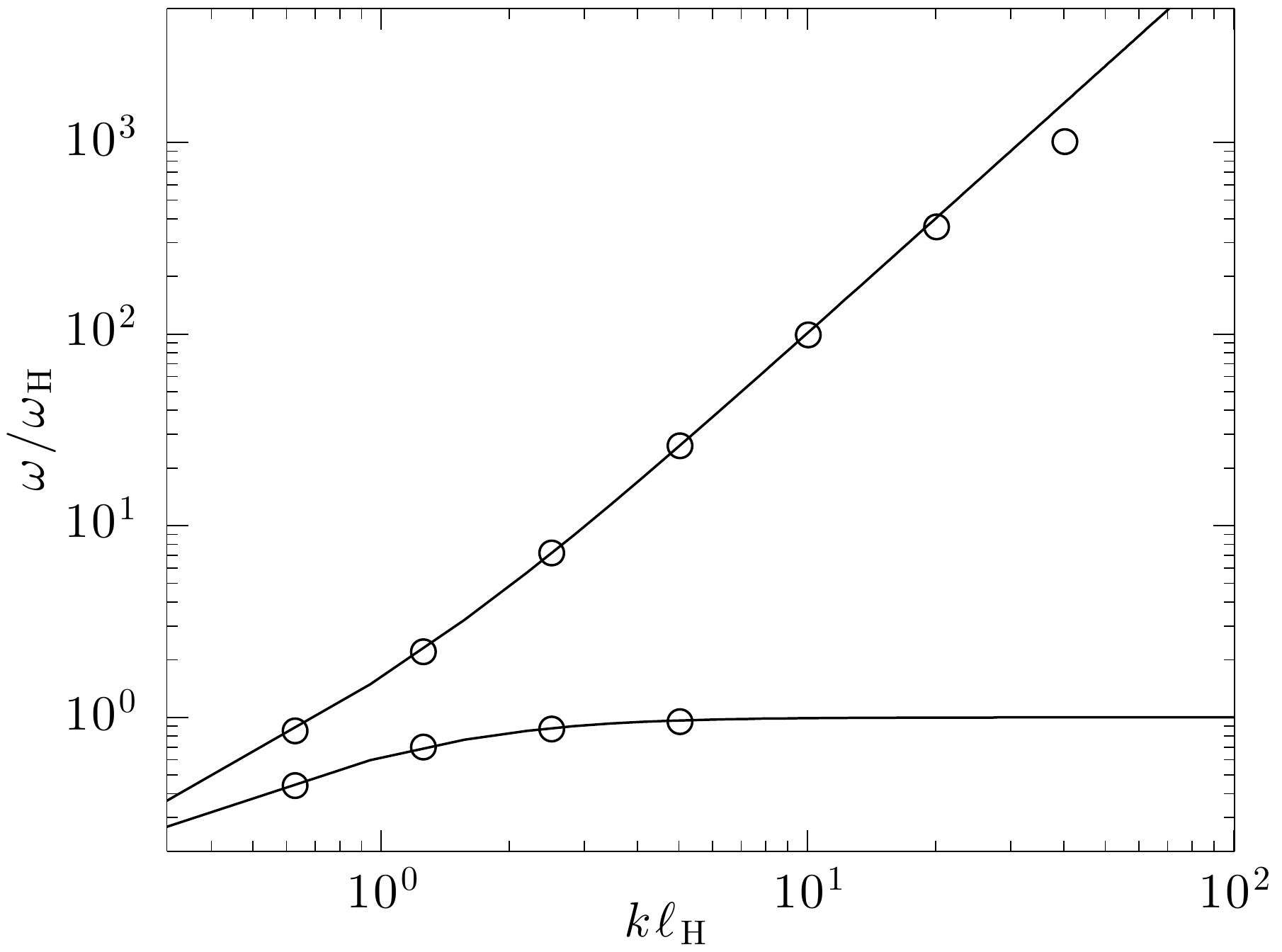}
\caption{Dispersion relation for whistler waves. Black line: analytical prediction; circles: eigenfrequencies measured 
in \textsc{Pluto} using our implementation of the Hall effect.}
\label{fig:numerical_whistlers}
\end{figure}

We test our integration scheme by first considering a simple configuration, one with a uniform mean magnetic field $\bm{B}
_0$ plus small sinusoidal perturbations $\delta\bm{B}_\perp$ perpendicular to $\bm{B}_0$. We then compute 
the frequency of the oscillation obtained in the code and compare it to the theoretical dispersion relation (see 
KL13 for details). The results are presented in Fig.~\ref{fig:numerical_whistlers}, where we have quantified the 
Hall effect by the Hall lengthscale $\ellH$ and the Hall frequency $\omega_\mathrm{H}\equiv v_\mathrm{A}/\ellH$.

In these numerical calculations, the Nyquist frequency is such that $k\ellH=80$. We find very good agreement for the whistler 
branch up to half of the Nyquist frequency. The non-propagating ion-cyclotron wave is not obtained at large $k$ because of the large 
numerical dissipation: the dissipation rate of this wave is faster than its oscillation period. 

%
%

%
%
\subsubsection{Convergence and numerical dissipation}

The question of convergence in Hall-MHD was raised by \cite{T08}. Since the whistler wave speed $\propto$$(dx)^{-1}$ and since 
numerical dissipation is roughly proportional to the wave speed, the convergence of the code at second order is not 
guaranteed in the presence of fast whistler waves. As was demonstrated by \cite{T08}, this problem can be solved using 
a \emph{symmetric} slope limiter in the reconstruction scheme, which naturally leads to a higher-order numerical 
diffusivity. In order to verify this, we present in Table \ref{tab:dissipation} the numerical damping rate $\gamma / \omega_\mathrm{H}$ for a 
whistler wave at $k\ellH=20$ as a function of the number of points per wavelength $n$. We find that both the 
monotonized centered and Van Leer slope limiters show second-order convergence as the resolution per 
wavelength is increased. This is not the case for the minmod limiter, where only first-order convergence is obtained. These results 
support the \cite{T08} argument and demonstrate that great care is needed when choosing the slope limiter in 
Hall-MHD. Unless otherwise stated, we always use the monotonized centered slope limiter.

\cite{T08} also demonstrated that this algorithm leads to numerical dissipation $\sim$$c_{\rm w} \bm{U}^{(4)} (\Delta x)^3$ where $U^{(n)}$ stands for the $n$\ts{th} derivative with respect to $x$. This implies that
\begin{itemize}
\item numerical dissipation damps whistler waves at the grid scale in one oscillation period. This explains why it is not possible to measure the numerical eigenfrequency in Section \ref{sec:num_disp} at the Nyquist frequency;
\item at a given resolution, the damping rate decreases as $k^{-4}$ (this has been verified by our numerical implementation). Hence, numerical dissipation decreases very rapidly as one goes to larger scales.
\end{itemize}

Finally, KL13 have shown that higher-order time-integration schemes are one way to guarantee the stable propagation of  whistler waves if one desires spectral accuracy without any numerical dissipation. Because of the numerical dissipation inherent to our algorithm -- a natural consequence of the Godunov flux as defined above -- the use of a higher-order time-integration scheme is not required in \textsc{Pluto}. We use an explicit second-order--accurate Runge-Kutta scheme unless otherwise stated.

%
%
\begin{table}[]
\centering
\setlength{\tabcolsep}{20pt}
\begin{tabular}{ccc}
\hline
slope limiter & $n=8$ & $n=16$\\
\hline
MC    &26 & 5.6\\
VL     & 29 & 7.7\\
MM   & 40 & 17\\
\hline
\end{tabular}
\caption{\label{tab:dissipation}Numerical damping rate $\gamma / \omega_\mathrm{H}$ of a whistler wave as a function of the number of points 
per wavelength $n$ using the monotonized centered (MC), Van Leer (VL), and minmod (MM) slope limiters.}
\end{table}

%
%
\subsubsection{Unstratified shearing box\label{sec:hall-unstrat}}

%
%
\begin{figure}
\centering
\includegraphics[width=0.9\linewidth]{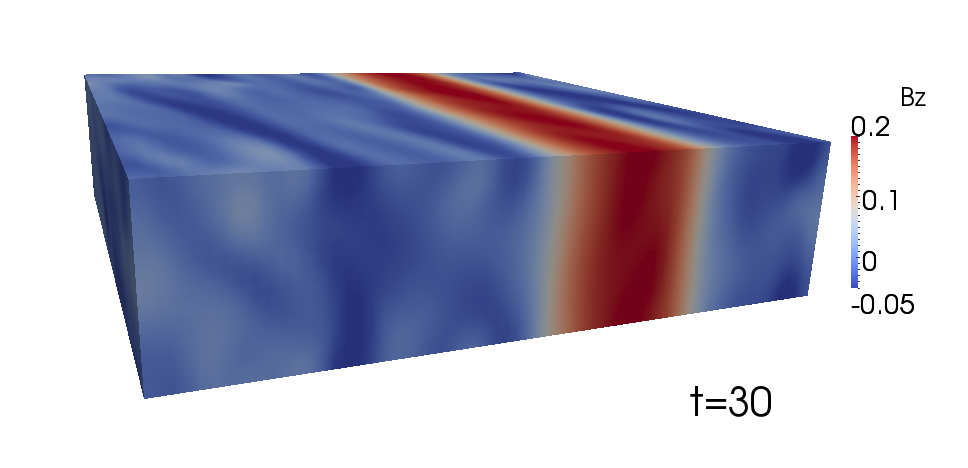}
\caption{Snapshot of $B_z$ in our unstratified Hall-MRI run at $t=30$. The flow exhibits a zonal-field structure similar to that discovered by KL13.}
\label{fig:hall_snap}
\end{figure}

KL13 demonstrated that the Hall-dominated MRI with $\langle B_y \rangle \lesssim \langle B_z \rangle$ 
saturates by producing large-scale axisymmetric structures in the magnetic field (``zonal fields''). 
To further test our Hall-MHD integration scheme and to control our numerical diffusivity, we have tried to reproduce this nonlinear behaviour in \textsc{Pluto}. 
We use very similar parameters to the run ZB1H1 of KL13. We use a $4\times 4\times 1$ box with a 
resolution $64\times 64\times 16$. We set $\ellH=0.55$, $B_{z0}=0.03$, and $\eta = \eta_{\rm A} = 0$. 
The resulting saturated state is shown in Fig.~\ref{fig:hall_snap} and 
exhibits a zonal-field structure similar to the one presented by KL13. This demonstrates that our 
scheme, despite being relatively diffusive due to the use of an HLL solver, manages to capture zonal-field 
structures with a resolution of 16 points per $H$.

%
%
\section{\label{app:ambi}Ambipolar diffusion}

%
%
\begin{figure}
\centering
\includegraphics[width=0.9\linewidth]{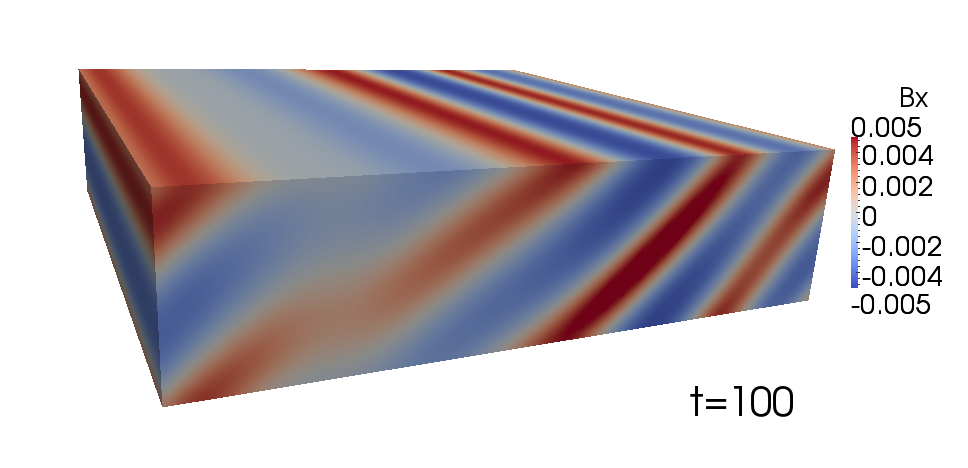}
\caption{Snapshot of $B_x$ in our ambipolar-MRI test simulation at $t=100$, just before saturation. The oblique nature of MRI channel mode, predicted by \citet{KB04}, is evident.}
\label{fig:ambi_snap}
\end{figure}

%
%
\begin{figure}
\centering
\includegraphics[width=0.9\linewidth]{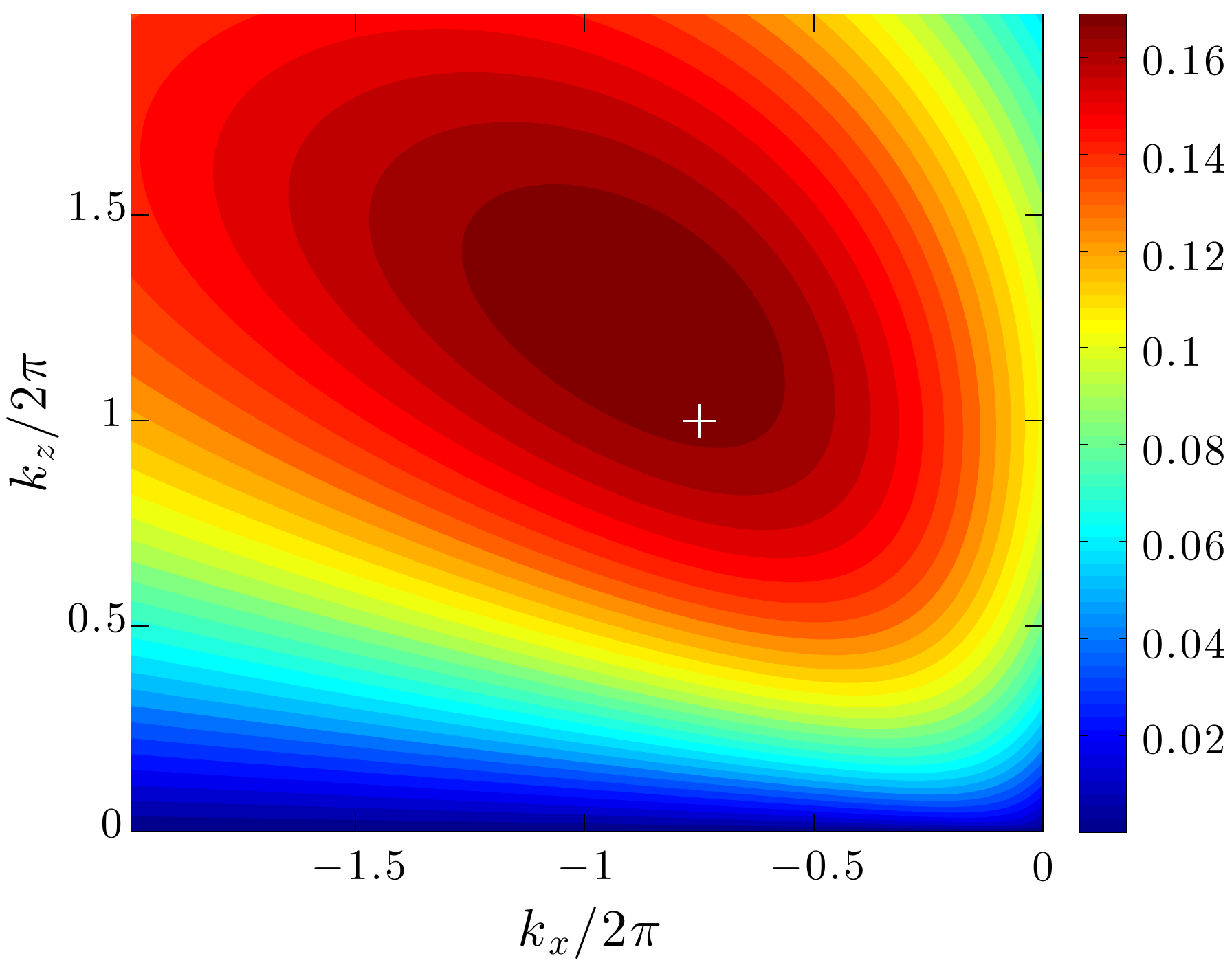}
\caption{Growth rate of the ambipolar-MRI for $B_{z0}=0.025$ and $B_{y0}=0.1$ as a function of the 
horizontal and vertical wavenumbers, $k_x$ and $k_z$. The largest growth rate available in the simulation obtains for $(k_x,k_z) = 2\pi \times (-0.75, 1)$ (white cross).}
\label{fig:ambi_growth}
\end{figure}

Ambipolar diffusion is implemented as a source term in a way analogous to that used for Ohmic resistivity. However, we noticed 
that grid-scale instabilities sometimes arise if the shearing-sheet boundary conditions are not enforced on the current density (a 
similar thing was observed with the Hall effect). Therefore, we use the current density computed in the conservative scheme for the 
Hall effect to compute the EMFs associated with ambipolar diffusion and Ohmic dissipation. This prevents small-scale instabilities from occurring 
at the radial boundaries.

We have tested our ambipolar diffusion module against the predictions of \cite{KB04}. An isothermal shearing box of size $4\times4\times 1$ and resolution $128\times 32\times 32$ is threaded by a mean magnetic field with both vertical and horizontal components, whose magnitudes are characterised by their respective Alfv\'{e}n speeds: $B_{z0}=0.025$ and $B_{y0}=0.1$. We introduce ambipolar diffusion by setting $\mathrm{Am}=1$. A typical snapshot of the linear growth phase is shown in Fig.~\ref{fig:ambi_snap}. We find the fastest-growing mode to have 
$(k_x, k_z) = 2\pi \times (-0.75,1)$. To check that this conforms to the expectation from linear theory, we show in Fig.~\ref{fig:ambi_growth} the 
theoretical growth rate of the ambipolar-dominated MRI as a function of $(k_x,k_z)$. We find that the mode seen in the 
the simulation corresponds to the fastest-growing mode from linear theory that is available in the simulation. The 
theoretical growth rate for this mode is $\gamma=0.171$; we find numerically $\gamma=0.17$. This demonstrates the 
accuracy of our implementation of ambipolar diffusion in \textsc{Pluto}.

\end{appendix}

\bibliographystyle{aa}
\bibliography{biblio}

\end{document}